\documentstyle[aps,preprint,epsfig]{revtex}
\tighten
\begin{document}

\def\mev{\hbox{\ MeV}}

\draft

%% Include following two lines for Journal Style

 %\twocolumn[\hsize\textwidth\columnwidth\hsize  %**Journal
 %\csname @twocolumnfalse\endcsname              %**Journal

\title{Branch points in the complex plane and \\  information loss
 in quantum systems at high level density}

\author{
I.~Rotter}
\address{
Max-Planck-Institut f\"ur Physik komplexer Systeme,
D-01187 Dresden, Germany}

\date{\today}

\maketitle

\vspace*{.5cm}

\begin{abstract}
The mechanism of  avoided level crossings in quantum systems is studied. It is
traced back to the existence of branch points in the complex plane which
influence the properties of resonance states as well as of discrete states. 
An avoided level crossing of two states causes not only an exchange of the two 
wave functions but, above all, correlations between them. The correlations  
play an important role at high level density since they cause the loss of 
information on the individual properties of the states. 

\end{abstract}

\vspace{.5cm}

\section{Introduction}

Recently, the generic properties of many-body
quantum systems are studied with a renewed
interest. They are expressed mostly by  their statistical 
properties, i.e. by comparing their level distributions with those
following from random matrix ensembles.  
The generic properties are, as a rule,
well expressed in the center of the spectra where the level density is high. 
As an example, the statistical properties of the shell-model states of nuclei 
around $^{24}$Mg  are
studied a few years ago \cite{zelev} by using two-body forces which are 
obtained  by fitting the low-lying states of  different nuclei
of the $2s-1d$ shell. In the center of the spectra,
the generic properties are well expressed  in spite 
of the two-body character of the forces used in the calculations.

A surprising result  is obtained recently  in performing shell-model
calculations for the same systems with random two-body forces.
In spite of the random character of the forces,
the regular properties of the low-lying states 
are described quite well \cite{bertsch} in these calculations.
Further studies \cite{frank,kaplan} proved the relevance of the 
results obtained and could explain in detail 
even the regular properties at the border of the spectra
obtained from random two-body forces \cite{drozdz}.
Thus, generic properties of the spectra may arise from two-body forces, 
while regular properties may be described by means of random forces. This
contra-intuitive result is, at present, not fully understood.

The spectra of microwave cavities are not determined by  two-body 
forces. Nevertheless, the calculated spectra are similar to those  from 
nuclear reactions \cite{ropepise}. They show deviations from the 
spectra obtained from random matrix theory as well as
similarities with them.
The theoretical results obtained  are  confirmed
by experimental studies \cite{perostba}. Most interesting are, 
however, the results 
of an analysis of the statistical properties of the
city transport in Cuernavaca (Mexico). It is shown in \cite{seba} that the
subsequent bus arrivals display probability distributions conforming to those
given by the unitary ensemble of random matrices.

By all these studies,  the question on the origin of the generic
properties of many-body
quantum systems, observed at high level density, is raised anew.
As a result of these studies, neither two-body forces nor quantum effects are
decisive. The study of the city transport in Mexico can be understood as an
equilibrium state of an interacting one-dimensional gas under the assumption
that the information contained in the positions of the individual gas
particles is minimized \cite{seba}. 
The minimum information is caused obviously by some information exchange 
with the environment. 

Quantum systems having discrete states seem to exist isolated from an
environment. It is therefore an interesting question  whether or not some
information exchange with the environment can, nevertheless, take place
and provide a certain loss of information on the individual properties of the 
states.  

The effect of avoided level crossing (Landau-Zener effect)
is known and studied theoretically as well as experimentally for many years. 
It is a quite general  property of the discrete states
of a quantum system whose energies  will never cross
when there is  a certain non-vanishing interaction between them.
Instead, they  avoid crossing in energy and
their wave functions are exchanged when
traced as a function of a certain tuning parameter.
The avoided level crossings are related to the existence of 
exceptional points \cite{hemuro}. The relation of these points
to geometrical phases is studied experimentally
\cite{dubb} as well as theoretically \cite{geom1,geom2} 
in a microwave resonator by deforming it cyclically.
The results show non-trivial phase changes when  degeneracies 
appear near to one another \cite{geom2}.
The influence of level crossings in the complex plane onto the
spectra of atoms is studied in \cite{solov1}. In this case, 
the crossings are called hidden crossings \cite{solov2}.    

Usually, it is assumed that avoided level crossings do not
introduce any  correlations between the wave functions of the states
as long as the system parameter is different from the critical one 
at which the two states avoided  crossing. 
Counter-examples have been found, however, in recent numerical studies of the 
spectra of microwave cavities \cite{ropepise,reichl}. The introduced
correlations are traced back to  branch points in the complex plane
\cite{ro2000} which are the crossing points of the states when
continued into the complex plane. 

If it is really a general phenomenon that
branch points in the complex plane introduce correlations between discrete
states of a closed quantum system, then an
information exchange with the environment would take place  also in quantum
systems. This fact could support the analogy 
of the statistical properties  of the city transport to those 
of quantum systems  and, above all, the interpretation  of the statistical
properties given in \cite{seba}.

It is the aim of the present paper to study in detail the correlations
in a quantum system which are introduced by branch points in the complex
plane. First, the properties of the  states of an open quantum system in the
very neighbourhood of a branch point will be studied. In such a case, the
$S$ matrix has a double pole and the interaction  between the two states is
maximum. The double pole coincides with the branch point in the complex plane.
It is shown further, that the  branch point influences the properties of
the resonance states not only when the physical
conditions for a double pole of the $S$ matrix are fulfilled
but also when the states avoid crossing. This result corresponds
to those obtained  for the realistic case of 
laser induced continuum structures in atoms 
\cite{marost}. The influence of the branch points
is accompanied, in any case, by an exchange of the  wave functions
of the states. 

The properties of resonance states in the neighbourhood of
an avoided crossing are traced in the further studies of the present paper. 
More exactly: starting from the
branch point where  two resonance  states with given widths 
$\gamma_1^{\rm cr}$ and
$\gamma_2^{\rm cr}$ cross, the properties of two other
states with $\gamma_i$ smaller or larger $\gamma_i^{\rm cr}$  are studied. 
For these two states,   the physical conditions for the formation of
a double pole of the $S$ matrix are not fulfilled.
Finally, the $\gamma_i$ are chosen to be zero. This study allows to compare
the properties of resonance states with finite lifetime with
those of discrete states under the influence of a branch point in the complex
plane.  The results show how the branch points in the complex plane continue 
analytically into the function space of discrete states. 

The paper is organized as follows.
In section 2, the mathematical properties of branch points in the complex
plane and their relation to avoided level crossings are sketched. 
In section 3, numerical results 
for states at a double pole of the $S$ matrix as well as for states 
with avoided  crossing
are given. The main point studied are the correlations of the wave
functions introduced by an avoided  level crossing.  
The results are discussed in the last section. They show that 
some information loss appears in quantum systems at high level density
under the influence of the branch points in the complex plane. 
The information contained in the positions and wave
functions of the individual
states of the system  is minimized in a similar manner as in the city
transport in Mexico.

\section{Branch points in the complex plane}

The avoided crossing of discrete levels is related to the existence of branch
points in the complex plane. In order to illustrate this relation, let
us study the properties of the complex two-by-two Hamiltonian matrix 
 \begin{eqnarray}
{\cal H} =
 \left(
\begin{array}{cc}
 e_1(a) & 0 \\
0  &   e_2(a)
\end{array}
\right) -
 \left(
\begin{array}{cc}
 \frac{i}{2}\gamma_1(a) &  \omega\\
 \omega &  \frac{i}{2} \gamma_2(a)
\end{array}
\right)  \; .
\label{eq:matr1}
\end{eqnarray}
The unperturbed energies $e_i$ and widths $\gamma_i$
($i=1,2$) of the two states  depend  on 
the parameter $a$ to be tuned in such a manner that  
the two states may cross in energy (and/or width) when $\omega = 0$.
The two states interact only via the non-diagonal matrix elements 
$\omega $ which  may be complex, in general.
The  imaginary part of $\omega $ arises from the residuum of the  
integral describing the interaction via the continuum of decay channels,
see e.g. \cite{ro91,mudiisro}. The real part contains both the 
direct interaction of the two states and the principal value integral
which appears from the interaction via the continuum.
In the following, we consider only real $\omega$ (and
$\gamma_i$ independent of $a$)  since we are interested 
in the influence of the branch points onto the avoided crossing
of discrete states. The influence of complex $\omega$ on the crossing of
resonance states will be considered in a forthcoming paper.

The eigenvalues of  ${\cal H}$     are
\begin{equation}
{\cal E}_{\pm} \equiv E_{\pm} - \frac{i}{2}  \Gamma_{\pm} =
\frac{\epsilon_1 + \epsilon_2}{2} \pm \frac{1}{2} \;
\sqrt{(\epsilon_1 - \epsilon_2)^2 + 4 \omega^2}
\label{eq:avoi}
\end{equation}
with $\epsilon_j \equiv e_j - \frac{i}{2} \; \gamma_j \; \; (j=1,2)$.
According to equation (\ref{eq:avoi}),  two interacting discrete states 
(with $\gamma_i = 0$) avoid
always crossing since $\omega$ is real in this case. It is, however, possible 
to find the corresponding crossing point in the complex plane by
considering non-vanishing values  $g_i$ instead of the $\gamma_i$.
Both, the $g_i$ and the $\gamma_i$, 
are proportional to the spectroscopic factors \cite{ro2000}. 
The common energy dependent factor in the $\gamma_i$ causes 
$\gamma_i \to 0$ in approaching the decay threshold where the resonance states
become discrete. The common energy dependent factor in the  $g_i$ is
different from zero also for discrete states. The  $g_i$ describe the transfer
probability between different states.

It can further be seen from 
equation (\ref{eq:avoi}) that
these crossing points are branch points in the complex plane. 
Equation (\ref{eq:avoi}) shows  also that  resonance states
with non-vanishing widths $\gamma_i$ avoid 
mostly crossing since
$  F(a,\omega) \equiv  (\epsilon_1 - \epsilon_2)^2 + 4 \omega^2 $
is different from zero for all $a$, as a rule. Only when  
$ F(a,\omega) = 0$ at $a=a^{\rm cr}$ (and $\omega=\omega^{\rm cr}$),
the states cross. In such a case, the $S$ matrix has a double pole,
see e.g. \cite{newton}. Note that the branch point is determined by 
the  values $(\omega )^2$ and $(\epsilon_1 - \epsilon_2)^2$
but not by the signs of these values.
According to equation (\ref{eq:avoi}),   it lies at $ X
\equiv (1/2) \{\epsilon_1(a^{\rm cr}) + \epsilon_2(a^{\rm cr})\} $.

The left and  right eigenfunctions, $\Phi^{\rm lt}_i$ and 
$\Phi^{\rm rt}_i$ ($i = \pm$), 
of a non-Hermitian matrix are different from one another. Since the
matrix  (\ref{eq:matr1})  is symmetrical, it follows 
\begin{equation}
\langle\Phi^*_i|\; {\cal H} = \langle \Phi^*_i |\; {\cal E}_i  \qquad 
{\rm and} \qquad
{\cal H} \; | \Phi_i \rangle =  {\cal E}_i\; | \Phi_i \rangle \; ,
\label{eq:orth5}
\end{equation}
see e.g. \cite{marost,mudiisro,pegoro}.
Therefore,  $\Phi^{\rm lt}_i=\Phi^{\rm rt \; *}_i \equiv \Phi^ *_i $. 
The eigenfunctions of  ${\cal H}$ 
can be orthonormalized according to 
\begin{equation}
\langle\Phi^{\rm lt}_i|\Phi^{\rm rt}_j\rangle =  \langle \Phi^*_i|\Phi_j
\rangle =\delta_{ij}
\label{eq:orth}
\end{equation}
where $\Phi^{\rm rt}_j \equiv \Phi_j$. 
Equation (\ref{eq:orth}) provides
the bi-orthogonality relations
\begin{eqnarray}
\langle \Phi_i|\Phi_i \rangle =
\Re( \langle \Phi_i|\Phi_i \rangle) 
= \langle \Phi_j|\Phi_j \rangle \;
 \;  & ; & \; \; A \equiv   \langle \Phi_i|\Phi_i \rangle
\ge 1
\nonumber \\
 \langle \Phi_i|\Phi_{j\ne i} \rangle  =  
i \; \Im (\langle \Phi_i|\Phi_{j\ne i} \rangle )  
 =  -\langle \Phi_{j\ne i}|\Phi_i \rangle \; \; & ; & \; \; B \equiv  
| \langle \Phi_i|\Phi_{j\ne i} \rangle| \ge 0 \; . 
  \label{eq:biorth}
\end{eqnarray}
At the double pole, $ |\langle \Phi_i|\Phi_i \rangle| \to \infty $ and 
$|\langle \Phi_i|\Phi_{j\ne i} \rangle| \to \infty $ but 
$ \langle \Phi^*_i|\Phi_j \rangle =\delta_{ij}$
according to equation (\ref{eq:orth}).
 Numerical examples for 
the values $ \langle \Phi_i|\Phi_j \rangle $ with $i=j$ as well as 
with $i\ne j$ 
can be found in \cite{mudiisro,pegoro,pepirose,slaving}.
Further, at the double pole the two wave
functions are related by $\Phi_i = \pm \, i \, \Phi_{j \ne i}$ \cite{marost}.

The eigenfunctions $\Phi_i$ can be represented in the set of
basic wave functions $\Phi_i^0$ of the  unperturbed matrix 
corresponding to  $\omega =0$,
\begin{equation}
\Phi_i=\sum b_{ij} \Phi_j^0 \; .
\label{eq:mix}
\end{equation}
 In the critical region of avoided crossing, the
eigenfunctions are mixed: $b_{ii}=b_{jj}$  and 
$b_{ij}=-b_{ji}$ for $i\ne j$. The $b_{ij}$ are normalized according to 
equation (\ref{eq:orth}).

\section{Numerical results}

\begin{figure}
\begin{minipage}[tl]{7.5cm}
\psfig{file=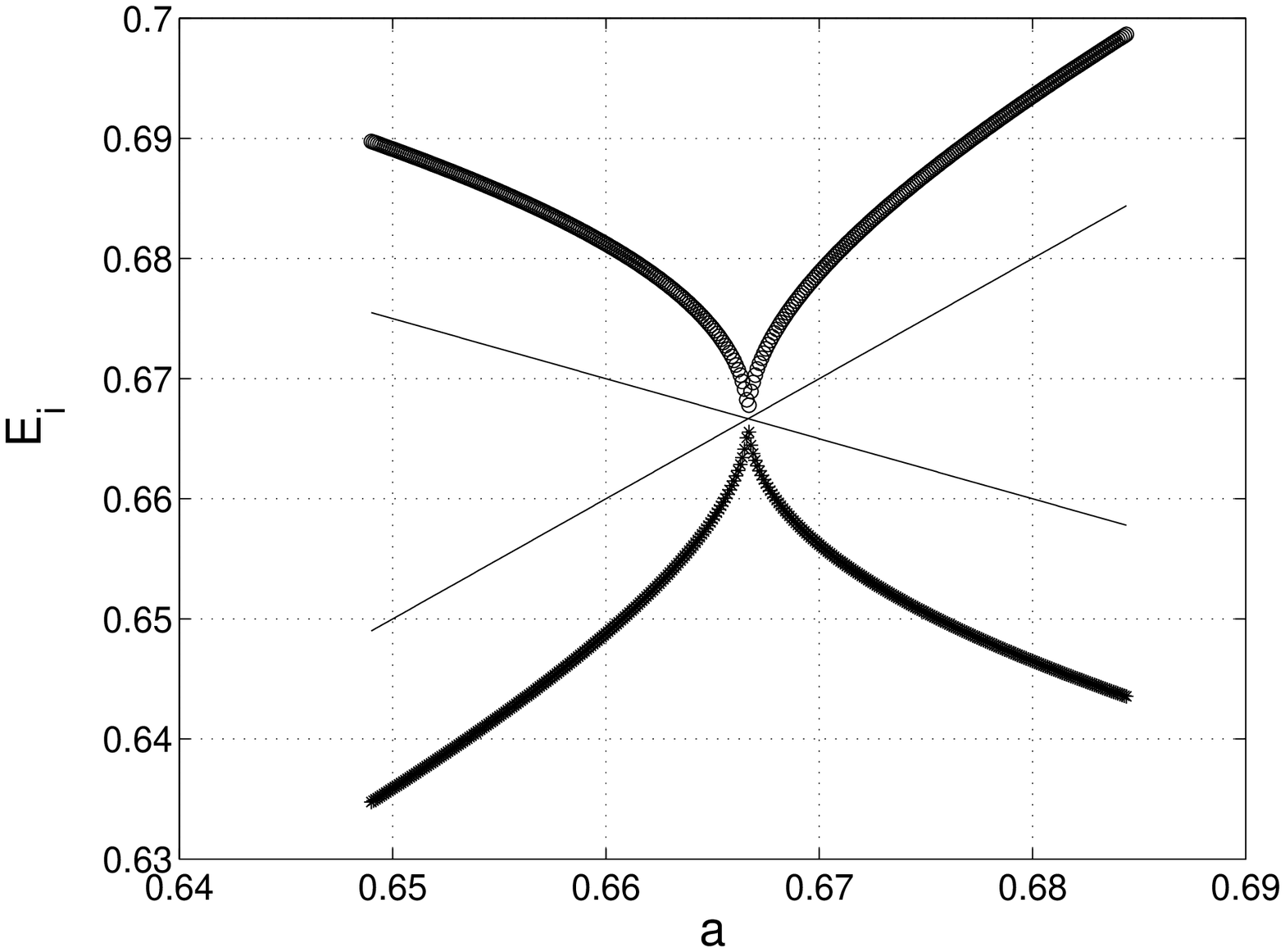,width=7.5cm}
\end{minipage}
\begin{minipage}[tr]{7.5cm}
\psfig{file=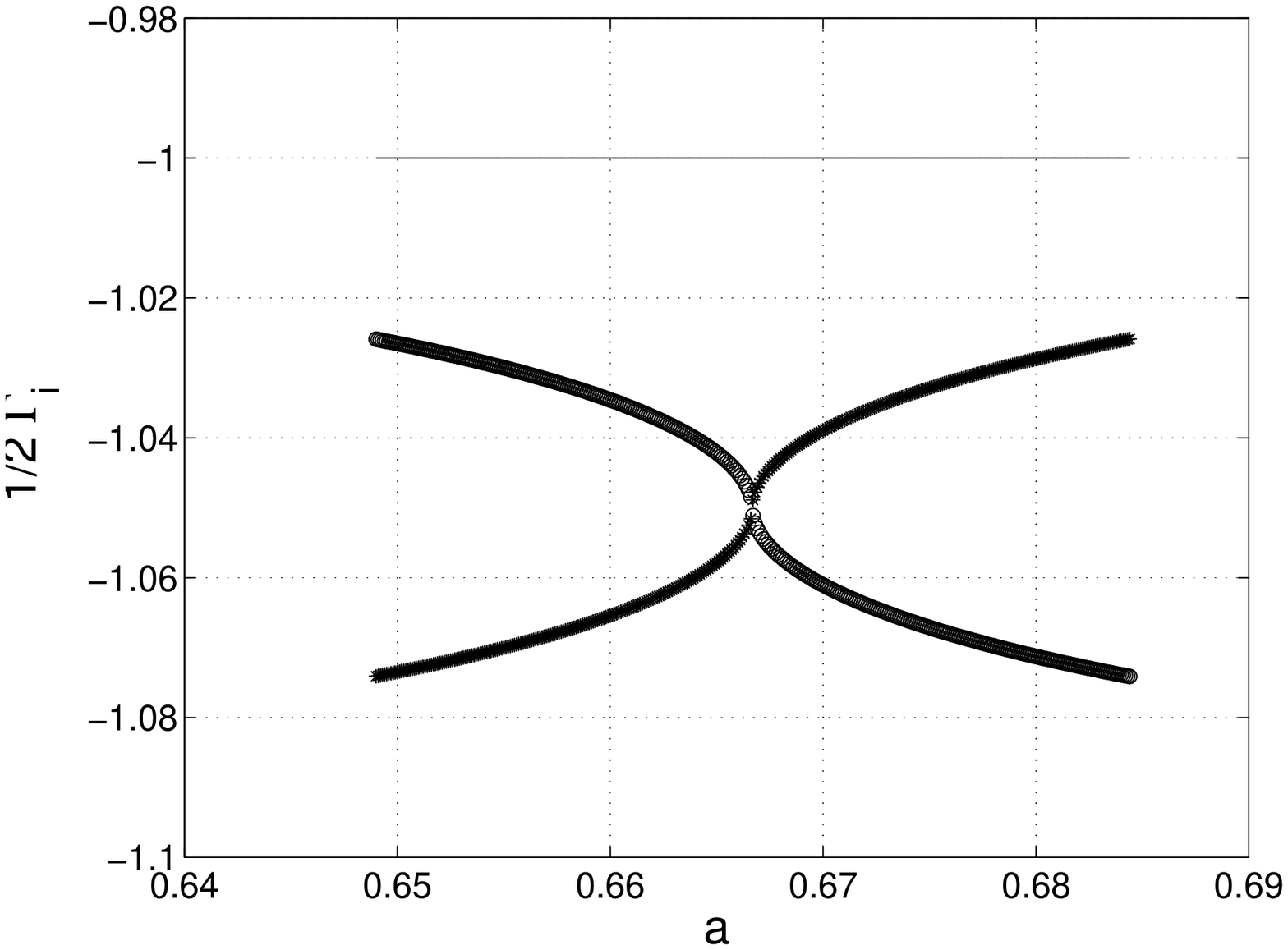,width=7.5cm}
\end{minipage}
\hspace*{.4cm}
\begin{minipage}[bl]{7.5cm}
\psfig{file=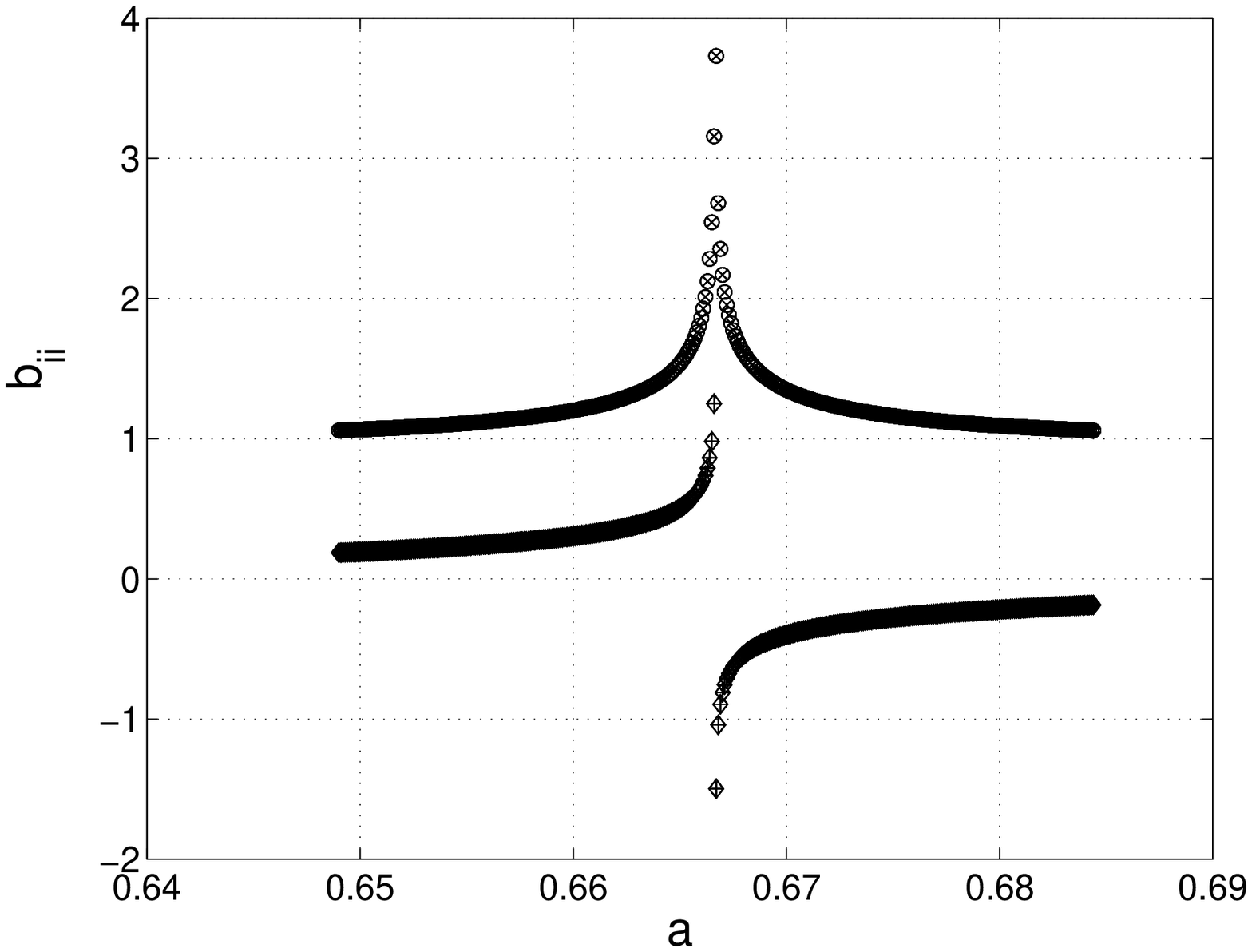,width=7.5cm}
\end{minipage}
\hspace*{.6cm}
\begin{minipage}[br]{7.5cm}
\psfig{file=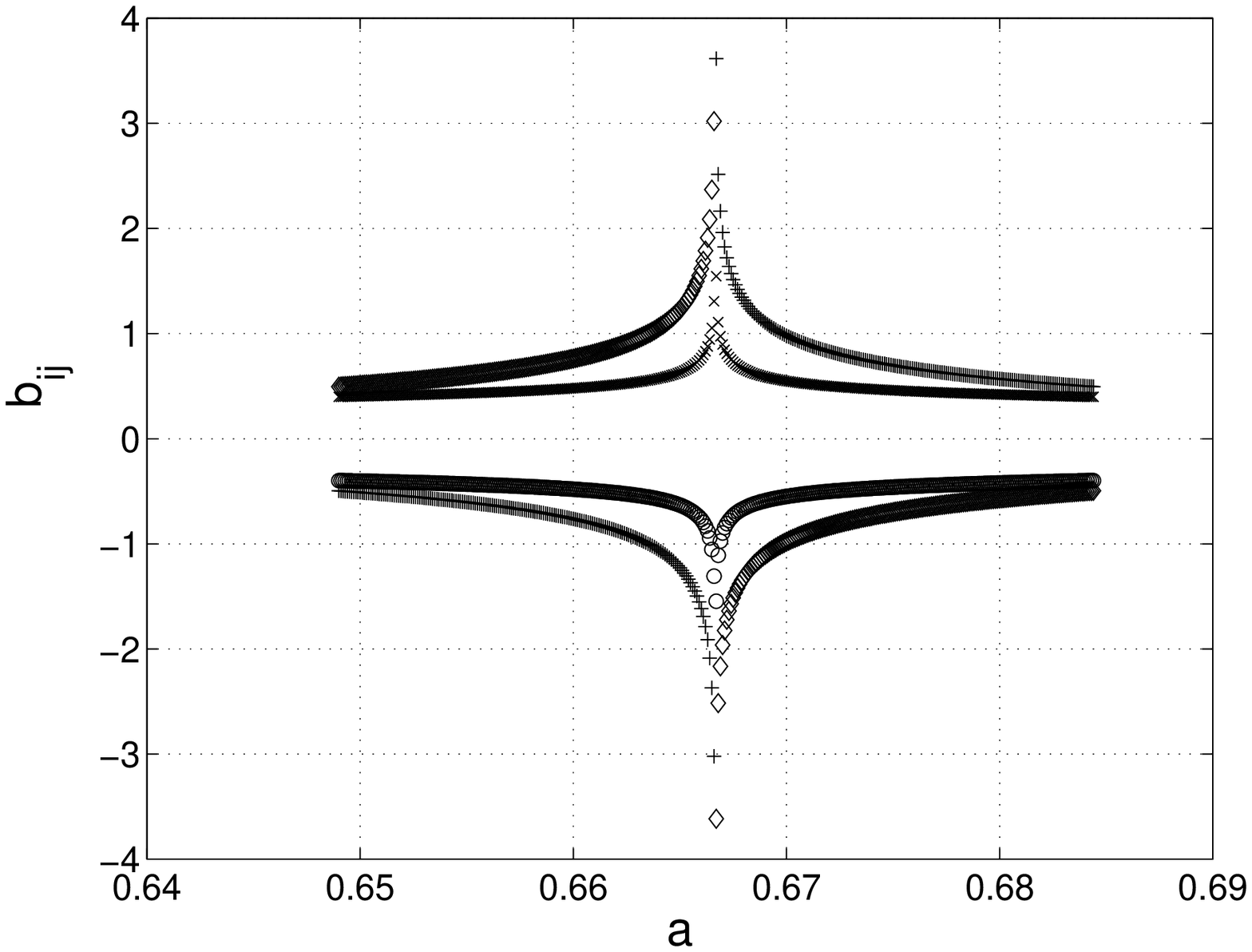,width=7.5cm}
\end{minipage}
\vspace*{.5cm}
\caption{The energies $E_i$ (top left) and widths
$\Gamma_i /2$  (top right) of   
the two eigenstates of the matrix (\ref{eq:matr1})
as a function of the parameter $a$. 
The thin lines give the energies $E_i$ and widths $\Gamma_i /2$ of the states 
at $\omega =0$.
The lower part of the figure shows the
 coefficients $b_{ii}$ (bottom left) 
 and  $b_{ij\ne i}$ (bottom right) defined by equation (\ref{eq:mix}). 
The  x and o denote 
the $\Re (b_{ij})$  while the  $\Im (b_{ij})$
are  denoted by + and $\diamond$. 
$e_1=1-a/2; \; e_2=a; \; \gamma_1 /2=1.0; \;
\gamma_2 /2 =  1.1$ and $\omega = 0.05$.
}
\label{fig:basic1}
\end{figure}

\begin{figure}
\hspace{-1.8cm}
\begin{minipage}[tl]{7.5cm}
\psfig{file=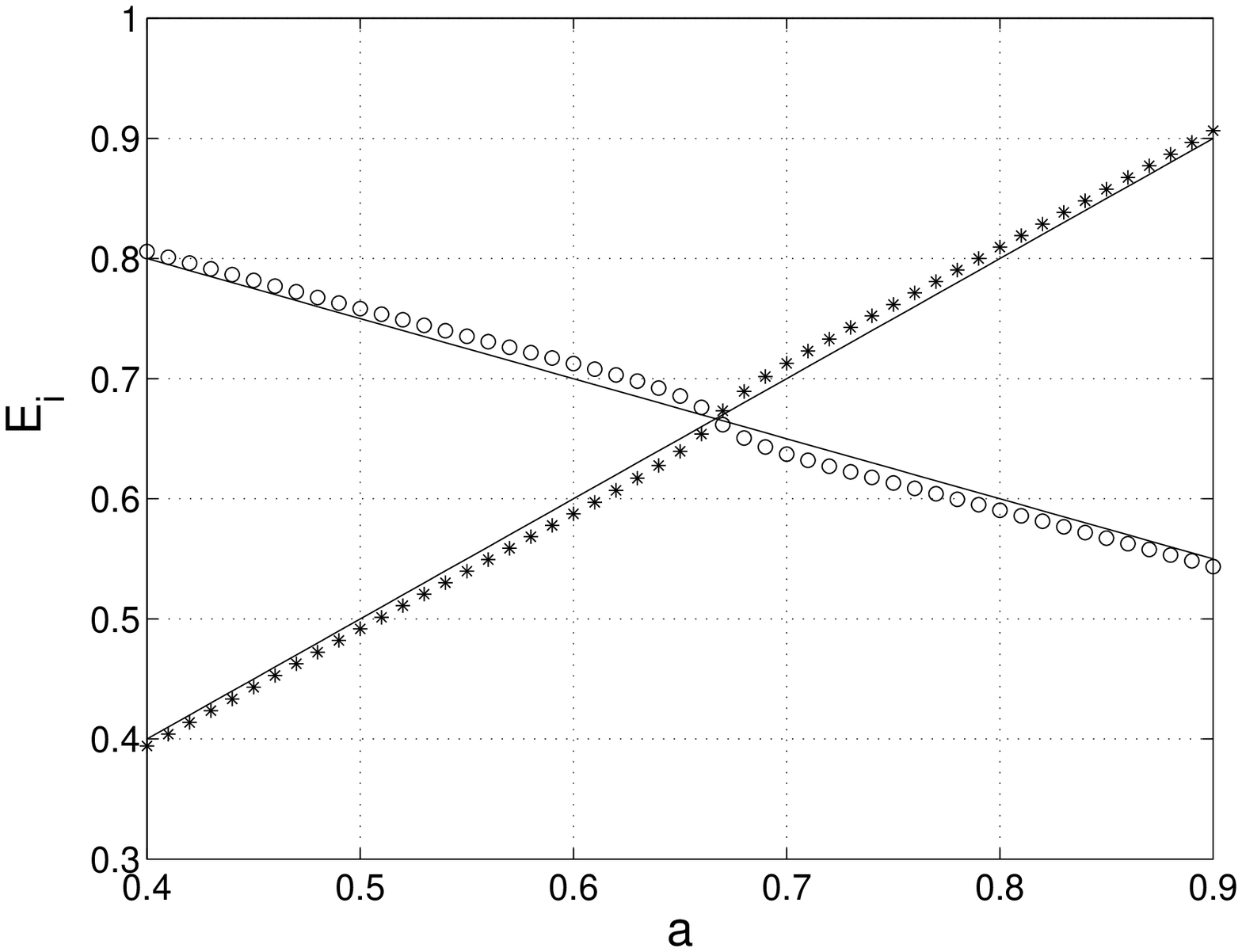,width=7.5cm}
\end{minipage}
\begin{minipage}[tr]{7.5cm}
\psfig{file=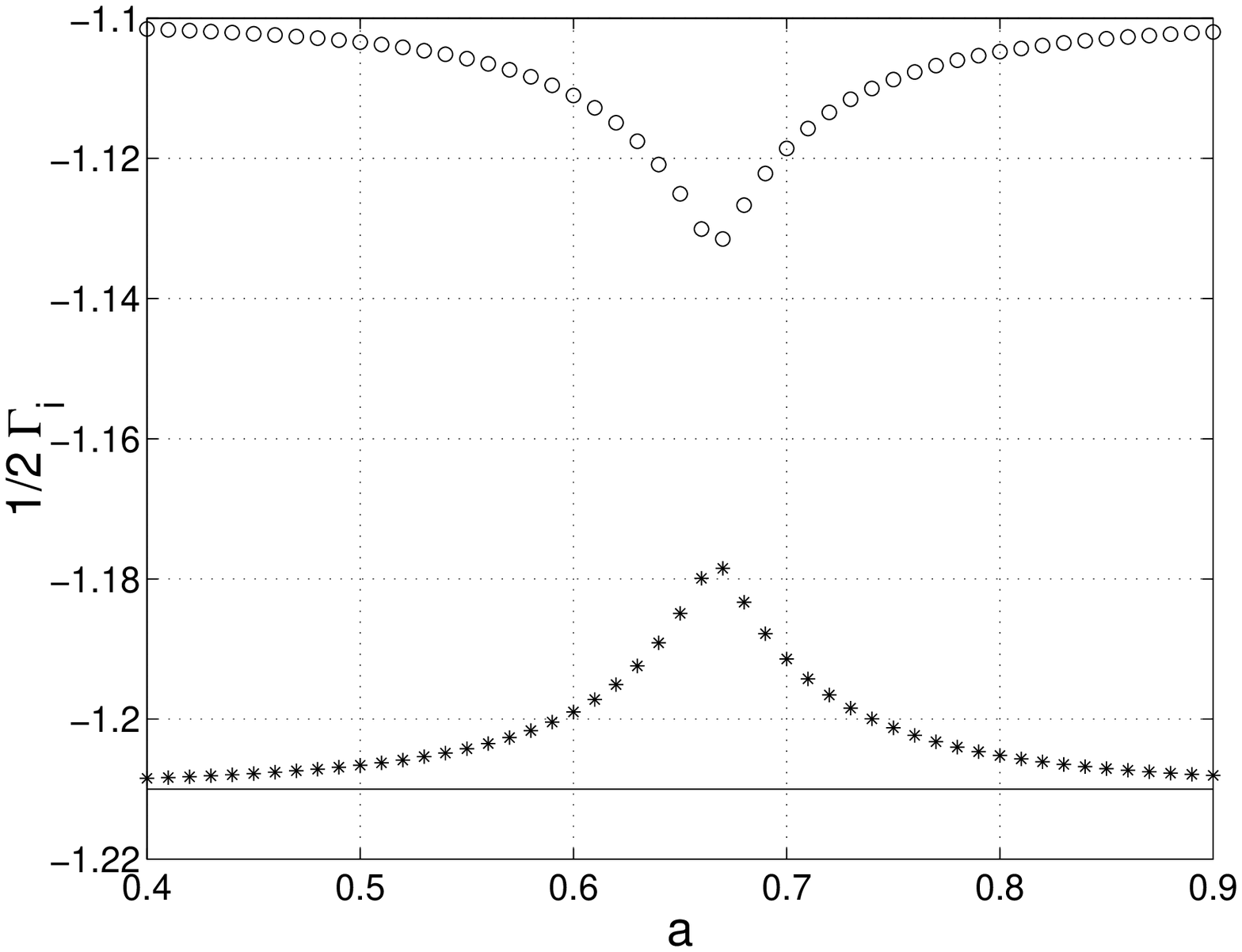,width=7.5cm}
\end{minipage}
\begin{minipage}[ml]{7.5cm}
\psfig{file=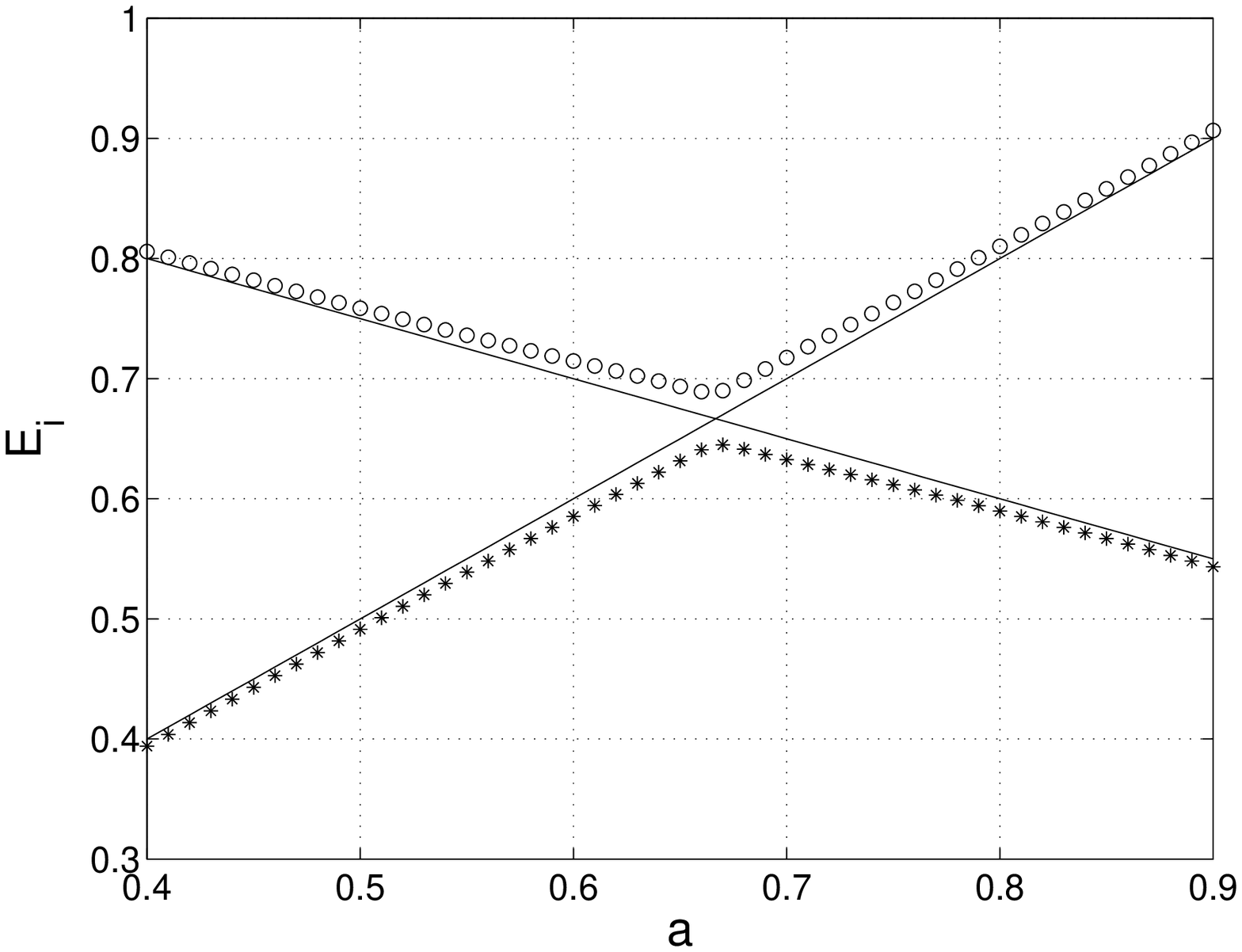,width=7.5cm}
\end{minipage}
\begin{minipage}[mr]{7.5cm}
\psfig{file=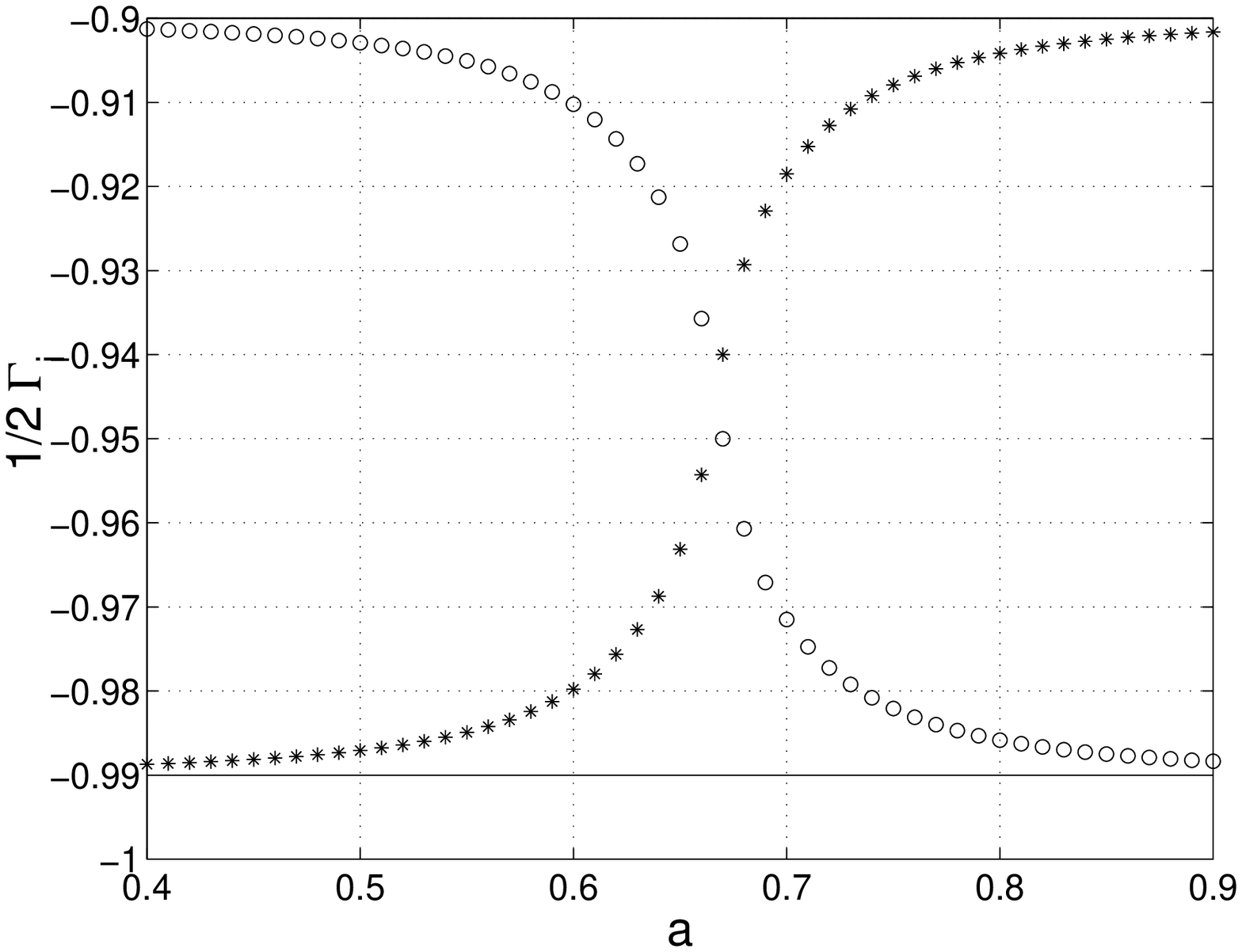,width=7.5cm}
\end{minipage}
\begin{minipage}[bl]{7.5cm}
\psfig{file=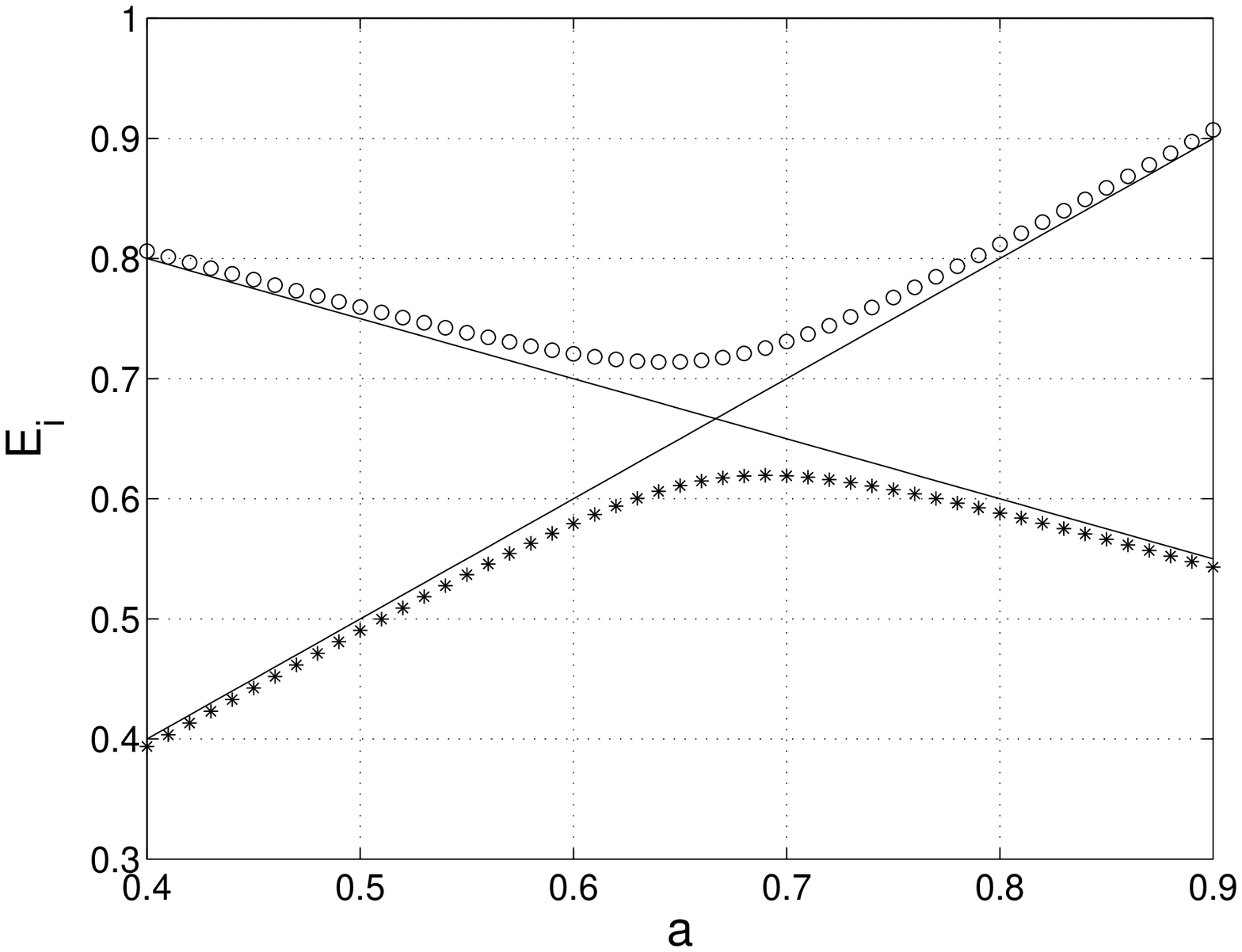,width=7.5cm}
\end{minipage}
\hspace{1.0cm}
\begin{minipage}[br]{7.5cm}
\psfig{file=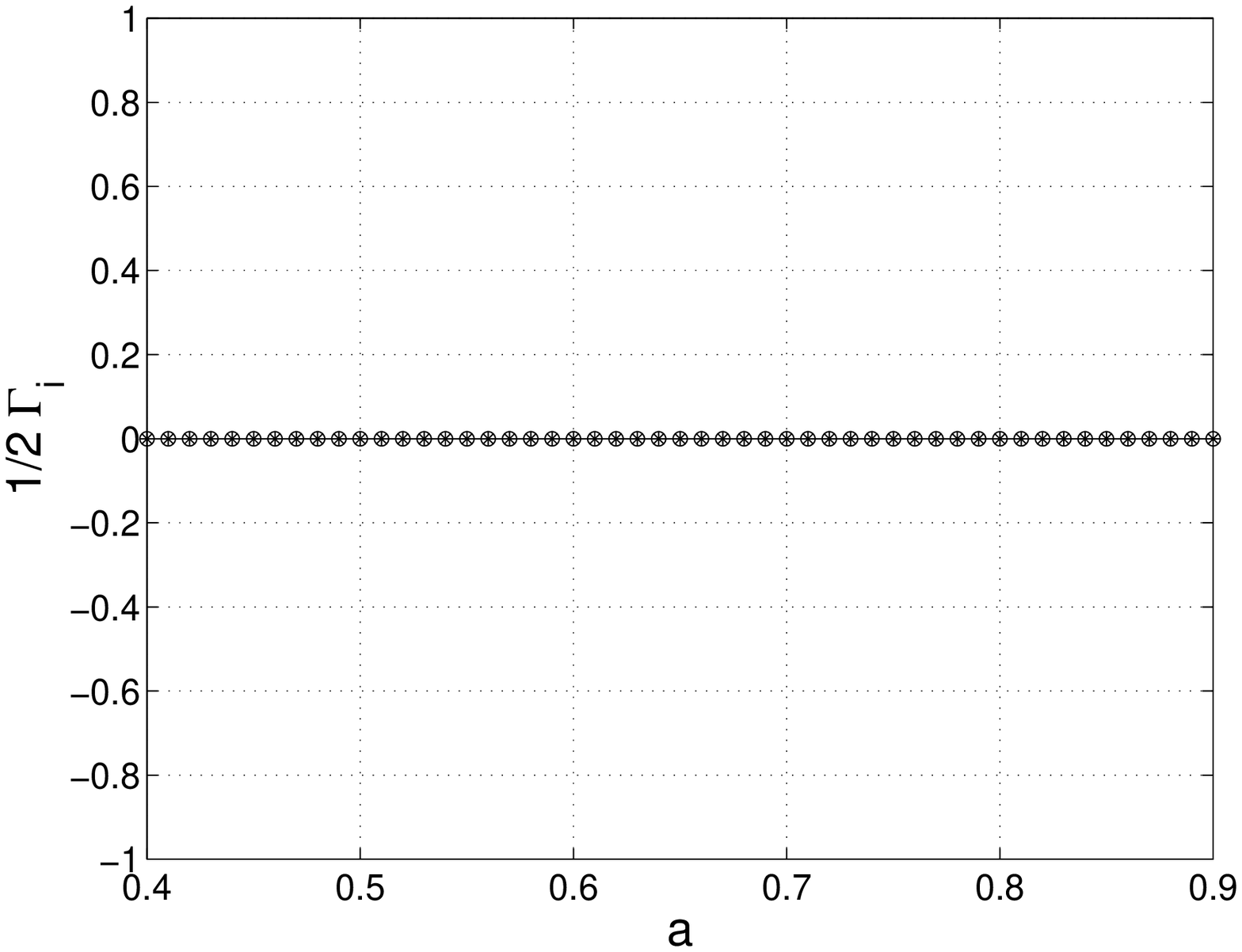,width=7.5cm}
\end{minipage}
\vspace*{.5cm}
\caption{The energies $E_i$ (left) and widths $\Gamma_i /2$
(right) as a function of the
tuning parameter $a$.  $e_1=1-a/2; \; e_2=a $ 
and $\omega = 0.05$. The 
$\gamma_1 /2 $ are 1.10  (top), 
0.90  (middle), 0  (bottom); $\gamma _2 = 1.1 \cdot \gamma _1$. 
The full lines show the $E_i$ and  $\Gamma_i /2$ for $\omega = 0$. 
}
\label{fig:tra1}
\end{figure}

\begin{figure}
\hspace{-1.8cm}
\begin{minipage}[tl]{7.5cm}
\psfig{file=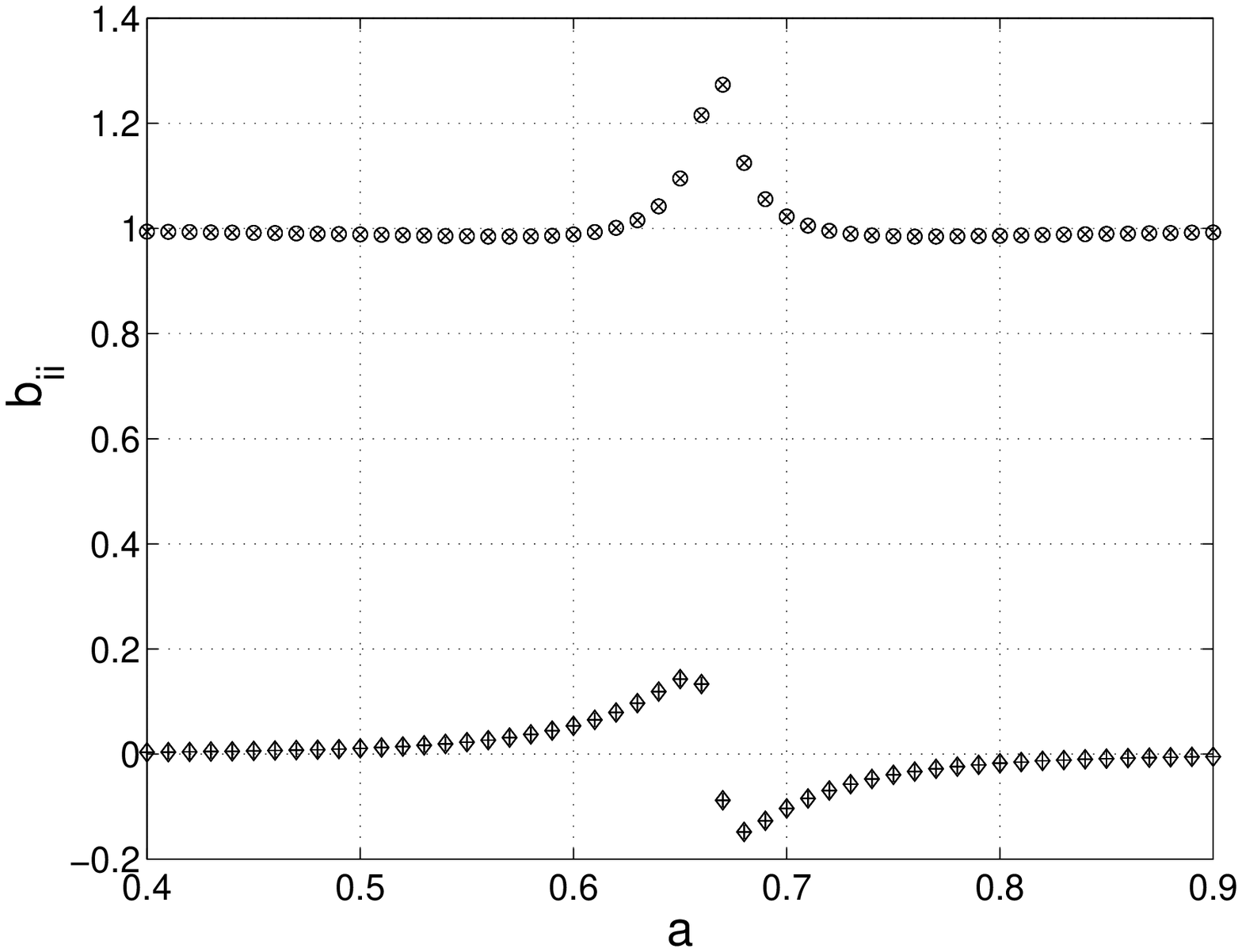,width=7.5cm}
\end{minipage}
\begin{minipage}[tr]{7.5cm}
\psfig{file=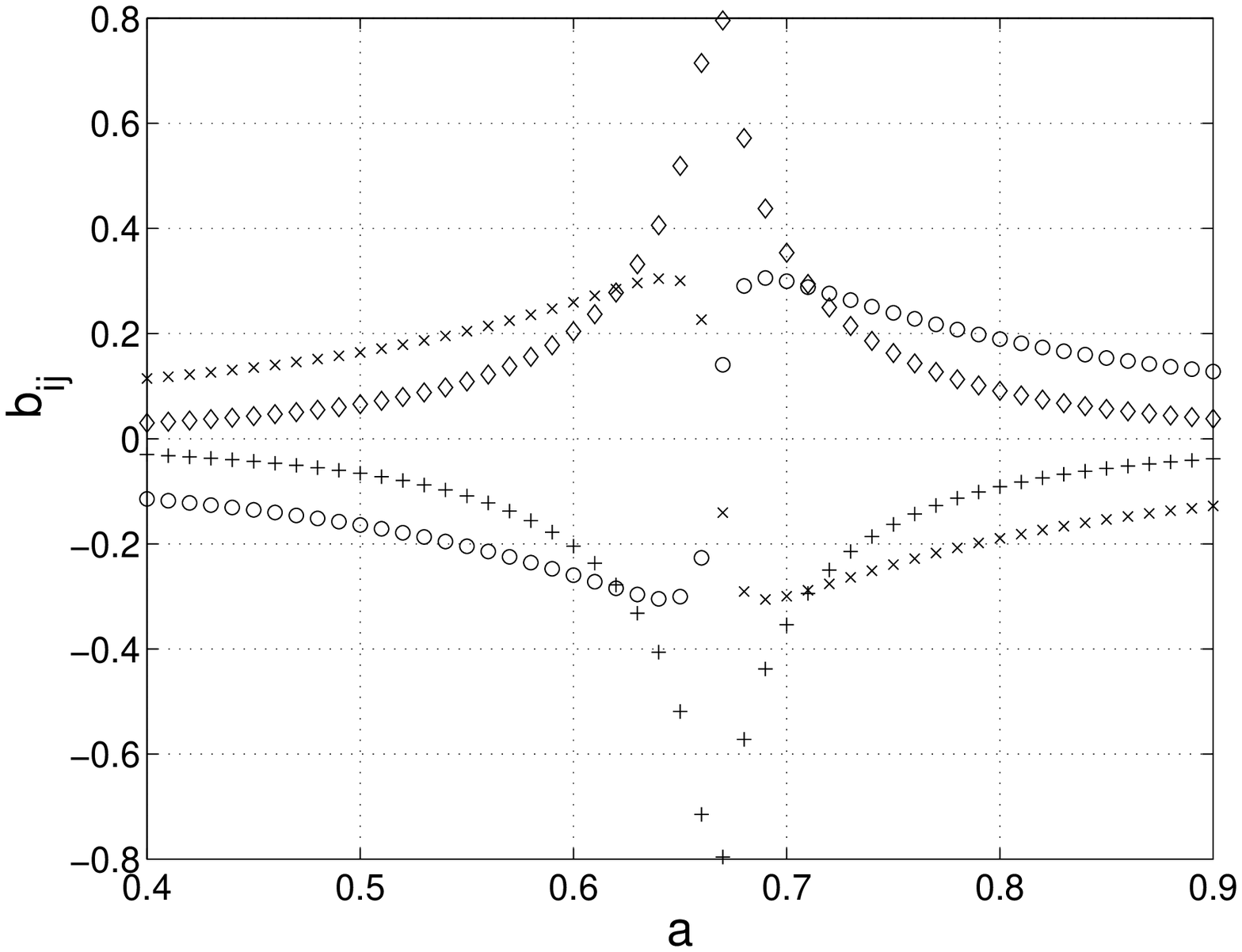,width=7.5cm}
\end{minipage}
\begin{minipage}[ml]{7.5cm}
\psfig{file=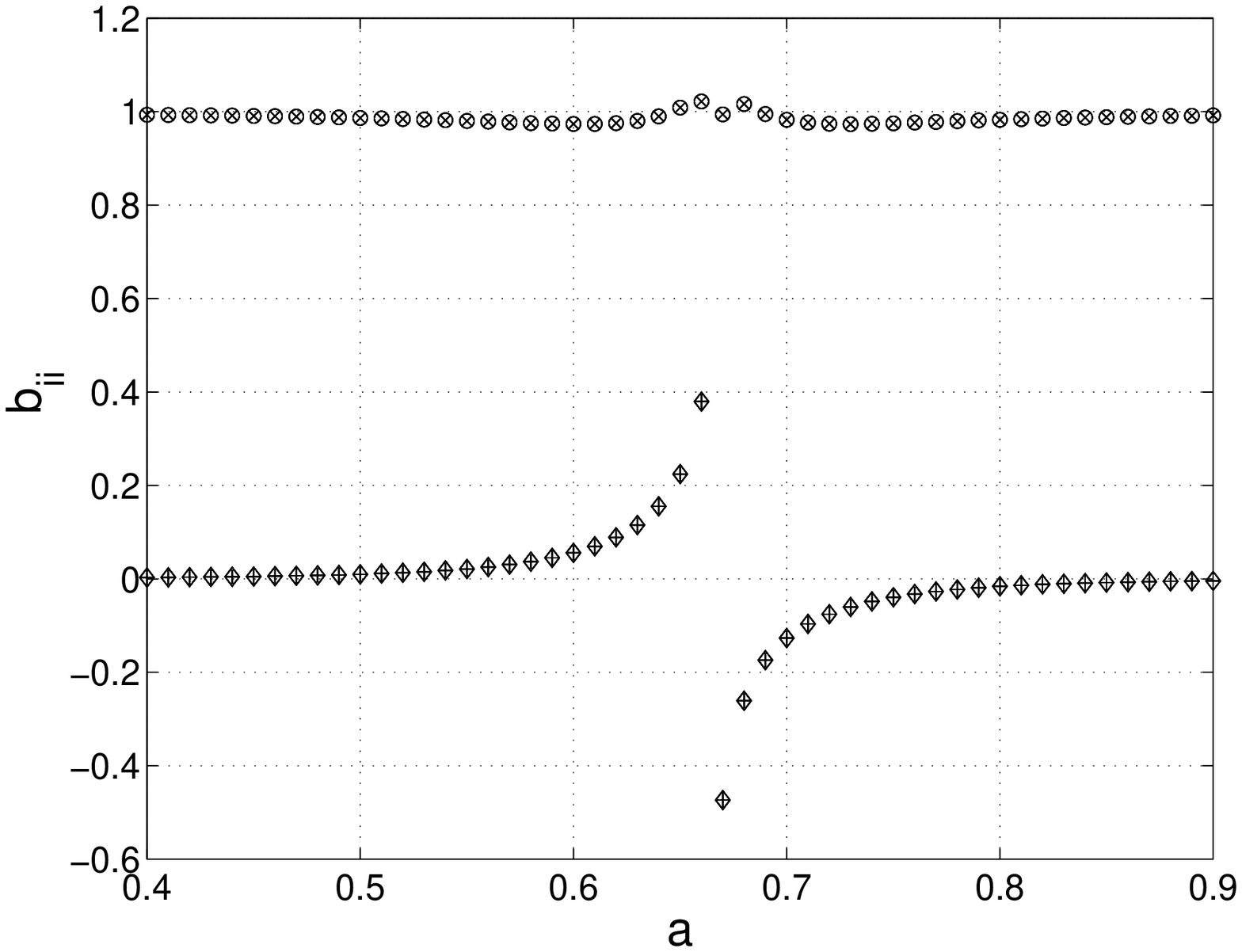,width=7.5cm}
\end{minipage}
\begin{minipage}[mr]{7.5cm}
\psfig{file=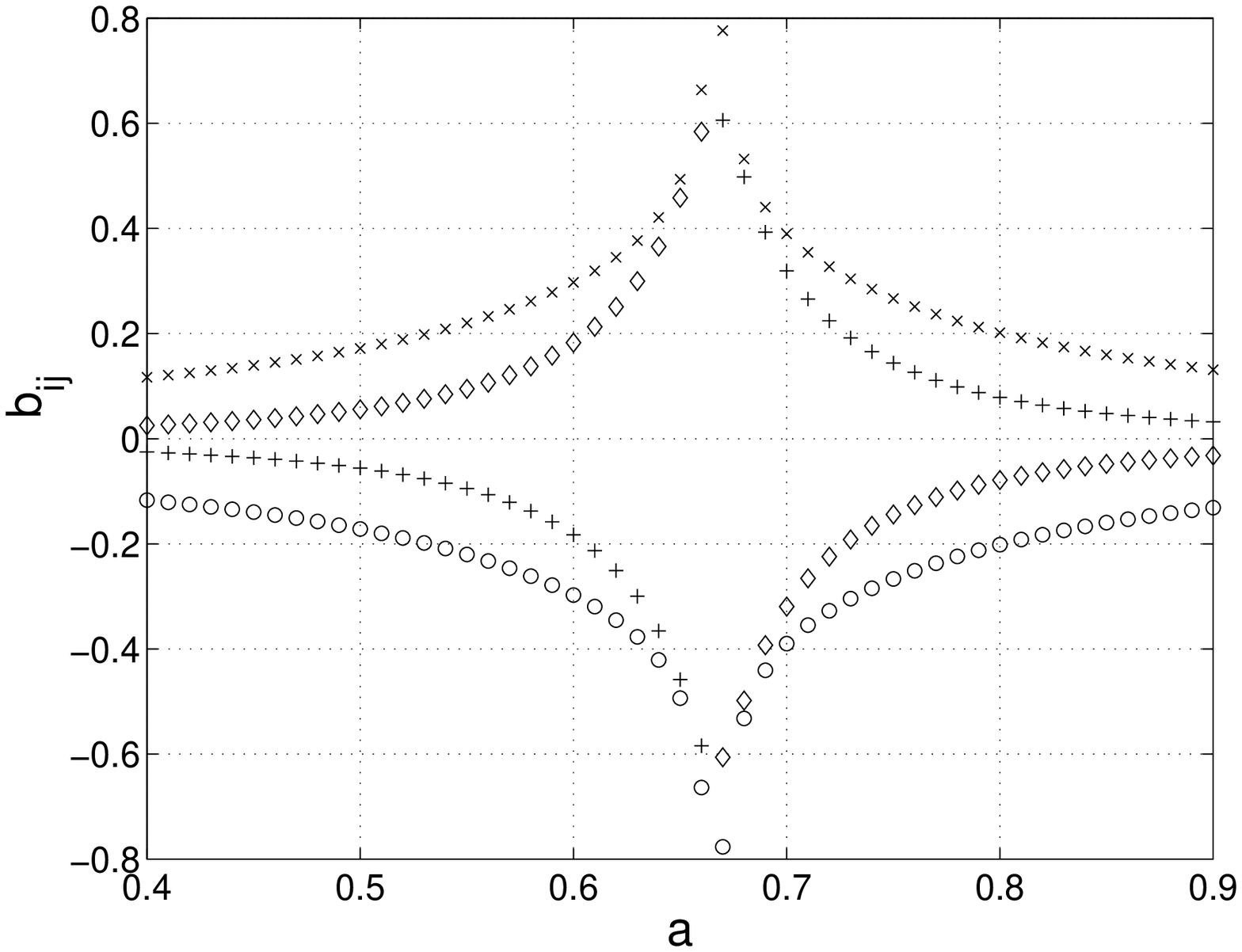,width=7.5cm}
\end{minipage}
\begin{minipage}[bl]{7.5cm}
\psfig{file=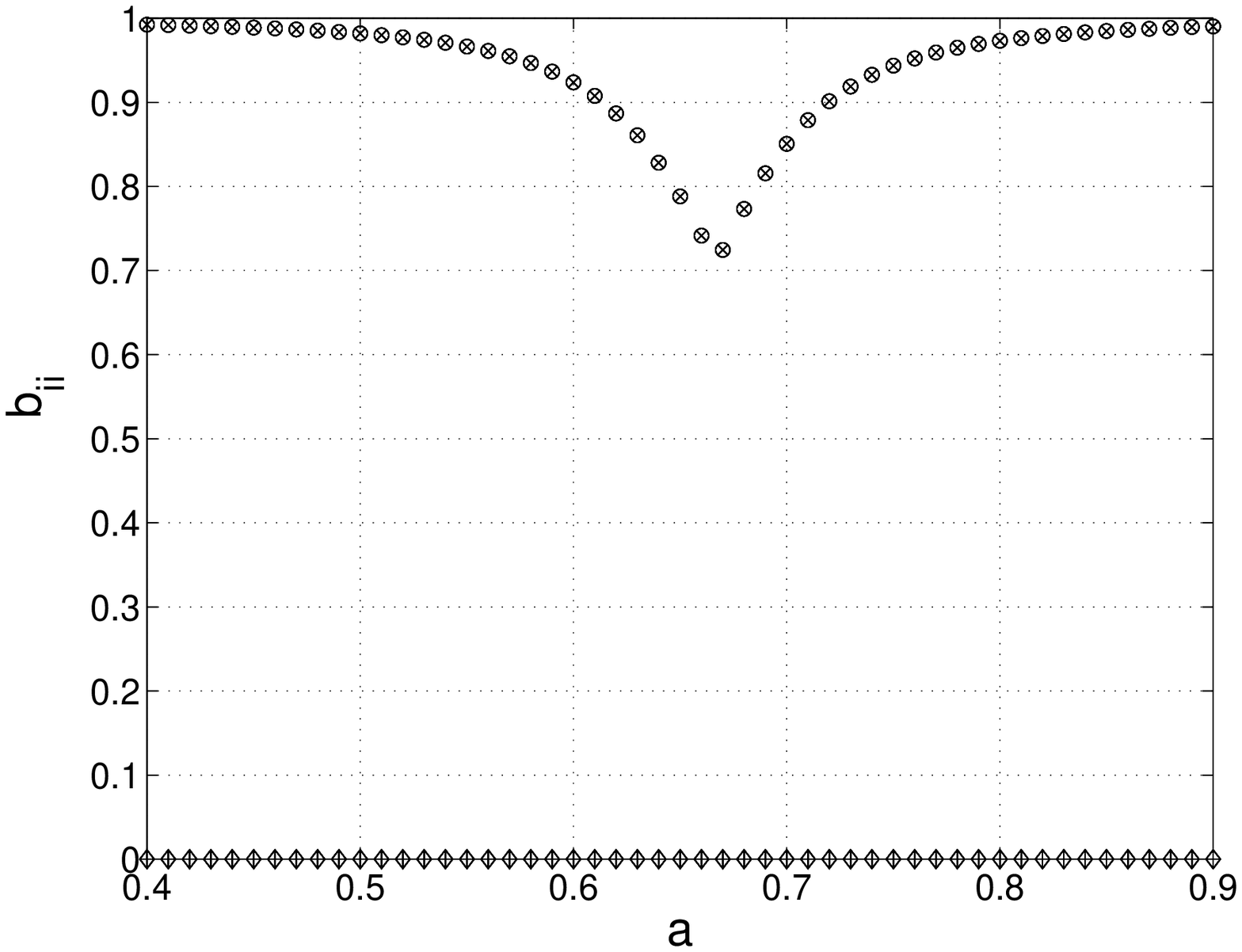,width=7.5cm}
\end{minipage}
\hspace{1.0cm}
\begin{minipage}[br]{7.5cm}
\psfig{file=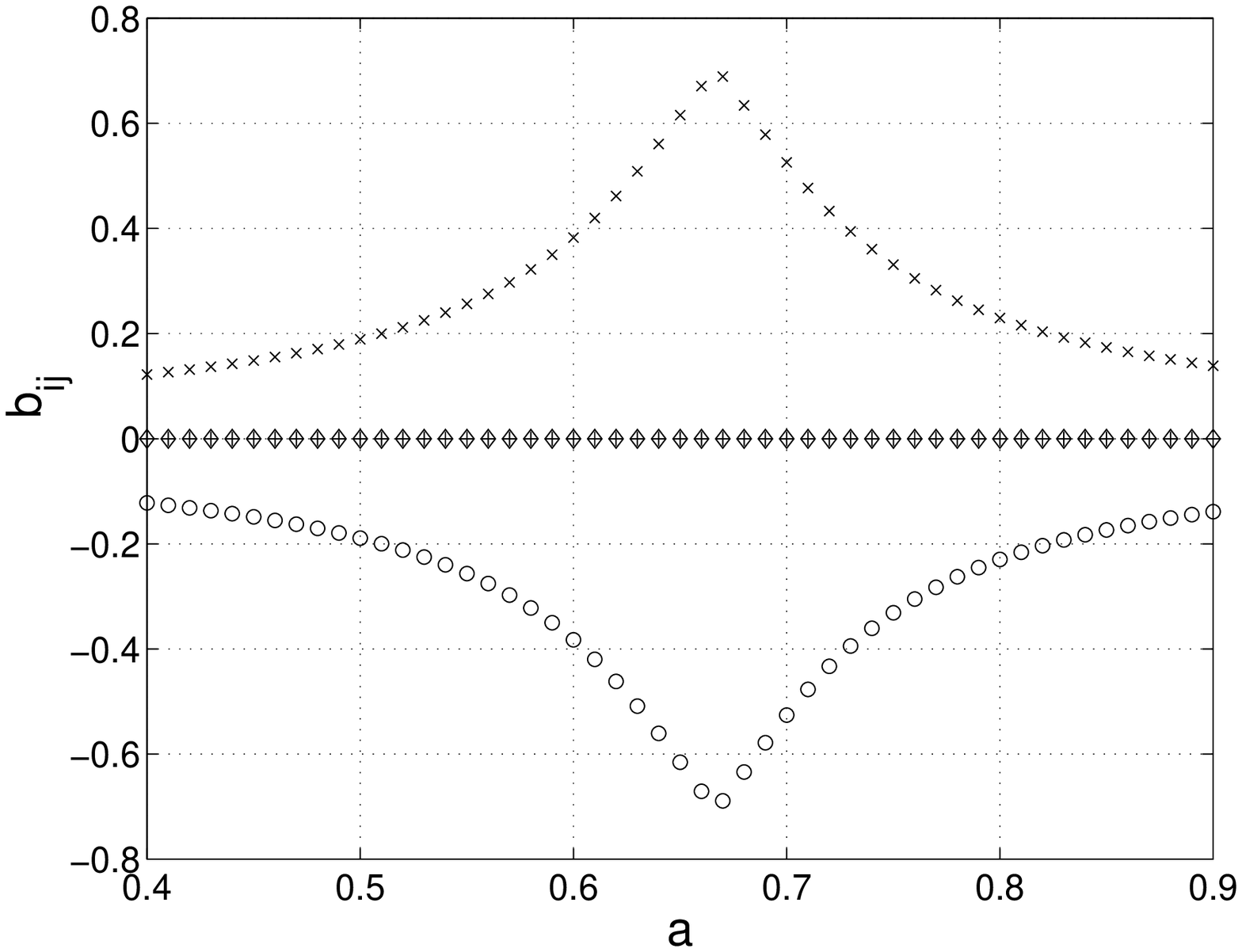,width=7.5cm}
\end{minipage}
\vspace*{.5cm}
\caption{The mixing coefficients $b_{ii}$ (left)
and $b_{ij\ne i}$ (right) defined by
equation (\ref{eq:mix}) as a function of the tuning parameter $a$.
o and x denote the real parts and $\diamond$ and + the imaginary parts. 
  $e_1=1-a/2; \; e_2=a $ and $\omega = 0.05$. The 
$\gamma_1 /2 $ are  the same as in figure \ref{fig:tra1}:
1.10  (top), 0.90  (middle), 0  (bottom);  $\gamma _2 = 1.1 \cdot \gamma _1$.
Note the different scales in the three cases.
}
\label{fig:tra3}
\end{figure}

The numerical results obtained by diagonalizing the matrix (\ref{eq:matr1})
are shown in figures \ref{fig:basic1} to \ref{fig:avoi3}.
In all cases $e_1=1-a/2, \; e_2=a $ and $\omega = 0.05$. The $\gamma_i$ do
not depend on the tuning parameter $a$.  At $a = a^{\rm cr}
=2/3$, the two levels cross when
unperturbed (i.e. $\omega = 0$) and avoid crossing, as a rule, when perturbed 
by the  interaction $\omega$. When $\gamma_1 /2 =
 \gamma_1^{\rm cr} /2 
=  1.0, \; \gamma_2 /2 = \gamma_2^{\rm cr} /2
= 1.1$, the $S$ matrix has a double pole meaning that the two resonance 
states cross in spite of $\omega \ne 0$. According to equation (\ref{eq:avoi}),
the double pole of the $S$ matrix is
a branch point in the complex plane. 

In figure \ref{fig:basic1}, the energies $E_\pm$, widths $\Gamma_\pm$ and wave
functions $b_{ij}$ of the two states are shown as a function of the parameter 
$a$ in the very neighbourhood of the branch point.   
Approaching the branch point at $a^{\rm cr}$, 
$\; |\Re (b_{ij})| \to \infty$  and $|\Im (b_{ij})| \to
\infty$. While  $\Re (b_{ij})$ does not change its sign by crossing the 
critical
value $a^{\rm cr}$, the phase of  $\Im (b_{ij})$ jumps from $\pm$ to $\mp$. 
The  orthogonality relations (\ref{eq:orth}) are fulfilled for all $a$
including the critical value $a^{\rm cr}$.

Figure \ref{fig:tra1} 
shows the energies $E_\pm$ and widths $\Gamma_\pm$ of the two states for
values of $\gamma_i$ just above and below the critical ones $\gamma_i^{\rm
cr}$ and for $\gamma_i = 0$, i.e. for discrete states.  When $\gamma_i >
\gamma_i^{\rm cr}$, the widths of the two states approach each other 
near $a^{\rm cr}$ but the width of one of the states remains always larger than
the width of the other one. 
The two states cross freely in energy, and the wave
functions are not
exchanged after crossing the critical value  $a^{\rm cr}$. 

The situation is
completely different when $\gamma_i <
\gamma_i^{\rm cr}$. In this case, the states avoid crossing in energy  
while their widths  cross freely. After crossing the critical value  $a^{\rm
cr}$, the wave functions of the two states are exchanged. An exchange  of the
wave functions takes place also in the case of discrete states 
($\gamma_i = 0$). This latter
result is well known as Landau-Zener effect. It is directly related to the 
branch point
in the complex plane at $a^{\rm cr}$ as can be seen from figure
\ref{fig:tra1}. 
 
The wave functions $b_{ij}$ are shown in figure  \ref{fig:tra3}.
The states are mixed (i.e.  $|b_{ii}| \ne 1$ and $b_{ij\ne i} \ne 0$)
 in all cases in the neighbourhood of   $a^{\rm cr}$.  
In the case without exchange of the wave functions, $\Re (b_{ij})$ as well as 
$\Im (b_{ij})$ behave smoothly at $a^{\rm cr}$ while this is true only for 
$\Re (b_{ij})$ in the case with exchange of the wave functions. In this case, 
$\Im (b_{ij})$ jumps 
from a certain finite value $y$ to $-y$ at $a^{\rm cr}$.
The $\Im (b_{ij})$ of discrete states are zero. Thus, a jump in the 
 $\Im (b_{ij})$ can not appear. The $\Re (b_{ij})$, however, show a dependence
on $a$ which is very similar to that of resonance states  with exchange
of the wave functions ($\gamma_i < \gamma_i^{\rm cr}$). 

In order to trace the influence of the branch point in the complex plane
onto the correlations of discrete states, the
differences  $\delta = |b_{ii}|^2 - |b_{ij\ne i}|^2$ and the values
 $|b_{ij}|^2$
are shown in figures \ref{fig:avoi1} and \ref{fig:avoi2} for different values
$\gamma_i$ from $\gamma_i > \gamma_i^{\rm cr}$ to $\gamma_i=0$.
Most interesting is the change 
of the value $\delta$ from 1 to 0 at  $\gamma_i^{\rm cr}$. The relation
$|b_{ii}|^2 = |b_{ij\ne i}|^2 $ at  $\gamma < \gamma_i^{\rm cr}$ is the 
result from interference processes. It holds also at  $\gamma_i =0$, 
i.e. for discrete states. In this case,  
$|b_{ii}|^2 = |b_{ij\ne i}|^2 = 0.5$ at $a^{\rm cr}$. 

The values $A$ and $B$ characterizing the bi-orthogonality of the two 
wave functions are shown in figure \ref{fig:avoi3} for the same values of 
$\gamma_i$ as in figures \ref{fig:avoi1} and \ref{fig:avoi2}.
The $A$ and $B$ are similar for $\gamma_i - \gamma_i^{\rm cr} = \pm \Delta$ 
as long as   $\Delta$ is small. They approach 
 $A \to 1$ and $ B \to 0$ for $\gamma_i \to 0$.

In figure  \ref{fig:four}, the energies $E_i$ and mixing coefficients 
 $|b_{ij}|^2$ are shown for illustration for four discrete states with three
 neighboured avoided crossings as a function of $a$. 
In analogy to (\ref{eq:matr1}), the matrix is 
 \begin{eqnarray}
{\cal H}^{(4)} =
 \left(
\begin{array}{cccc}
 e_1(a) & 0 & 0 & 0\\
0  &   e_2(a) & 0 & 0 \\
0  &  0 & e_3(a) & 0 \\ 
0  & 0 & 0 &  e_4(a) 
\end{array}
\right) -
 \left(
\begin{array}{cccc}
 0 &  \omega_{12} & \omega_{13} & \omega_{14} \\
  \omega_{21} & 0 & \omega_{23} & \omega_{24} \\
  \omega_{31} & \omega_{32} & 0 & \omega_{34} \\
 \omega_{41} & \omega_{42} &  \omega_{43} & 0\\
\end{array}
\right)  \; .
\label{eq:matr4}
\end{eqnarray}
The correlations 
introduced by the avoided
 crossings remain  at all values of the parameter $a$  
at high level density.  They are the result of
 complicated interference processes. This can be seen best by comparing
 the two pictures with four interacting states (top and middle 
in figure  \ref{fig:four}) with those of only two
 interacting states (bottom of figure  \ref{fig:four} and bottom right of
 figure \ref{fig:avoi2}). Figure \ref{fig:four} bottom shows the large region
of the $a$ values around $a^{\rm cr}$ for which the two wave functions remain
mixed: $|b_{ii}|^2 \to 1$ and $|b_{ij\ne i}|^2 \to 0$ for $a \to a^l$ with 
$|a^l - a^{\rm cr}_{34}| \gg |a^{\rm cr}_{i4} - a^{\rm cr}_{34}| \; ; \; 
i=1,2$.
The avoided crossings between neighboured states do, therefore, not occur
between states with pure wave functions and it is impossible  to
identify the  $|b_{ij}|^2$ unequivocally (figure   \ref{fig:four} right top and
middle).

\section{Discussion of the results}

Most  calculations  represented in the present paper 
are performed for two states which cross or avoid crossing
under the influence of an interaction $\omega$ which is real. 
A general feature appearing in all the results is the  repulsion
of the levels in energy (except in the very neighbourhood of
$ a^{\rm cr}$ when $
\gamma_i \ge \gamma_i^{\rm cr}$). This result follows analyically from the
eigenvalue equation  (\ref{eq:avoi}). It holds quite generally for real
$\omega$  as discussed by means of
the spectra of microwave cavities   \cite{ropepise}
and laser-induced continuum structures in atoms \cite{marost}.
The level repulsion in energy is accompanied by an approaching of  the
lifetimes (widths)  of the  states. 

The value $a^{\rm cr}$ is determined by the free crossing in energy 
of the two non-interacting states $\Phi_i^0$ (corresponding
to $\omega = 0$) in the cases considered.
The avoided level crossing at the
critical value $a^{\rm cr}$ can be seen clearly only when the interaction
$\omega$ is small. For larger $\omega$, the avoided crossings are washed out
and are difficult to identify without performing a detailed analysis of 
the data. 
As a consequence, there are much more avoided level crossings in
quantum systems at high level density than usually believed and the 
influence of the branch points may be quite important. The last point does
not coincide with any intuitive assumptions.

The value of  the (real) interaction $\omega$ in relation to the value of the
 widths $\gamma_i$ is decisive whether or not the states will be exchanged at
 the critical value   $a^{\rm cr}$
of the tuning parameter, i.e. whether or not the states avoid crossing.
 When  $\omega$ is so small that the
 difference of the  widths $\Gamma_+ - \Gamma_-$ is different from zero at 
 $a^{\rm cr}$ then the states will not be exchanged. If, however, 
$\Gamma_+ = \Gamma_-$ at $a^{\rm cr}$, the states will be exchanged.
The exchange occurs due to interferences between the different components of
 the wave functions. This can be seen best for resonance states with  
$\gamma_i \ne 0$ where interferences between the real and imaginary parts of
 the wave functions play an important role (see figure
 \ref{fig:avoi2}).

\begin{figure}
\begin{minipage}[tl]{7.5cm}
\psfig{file=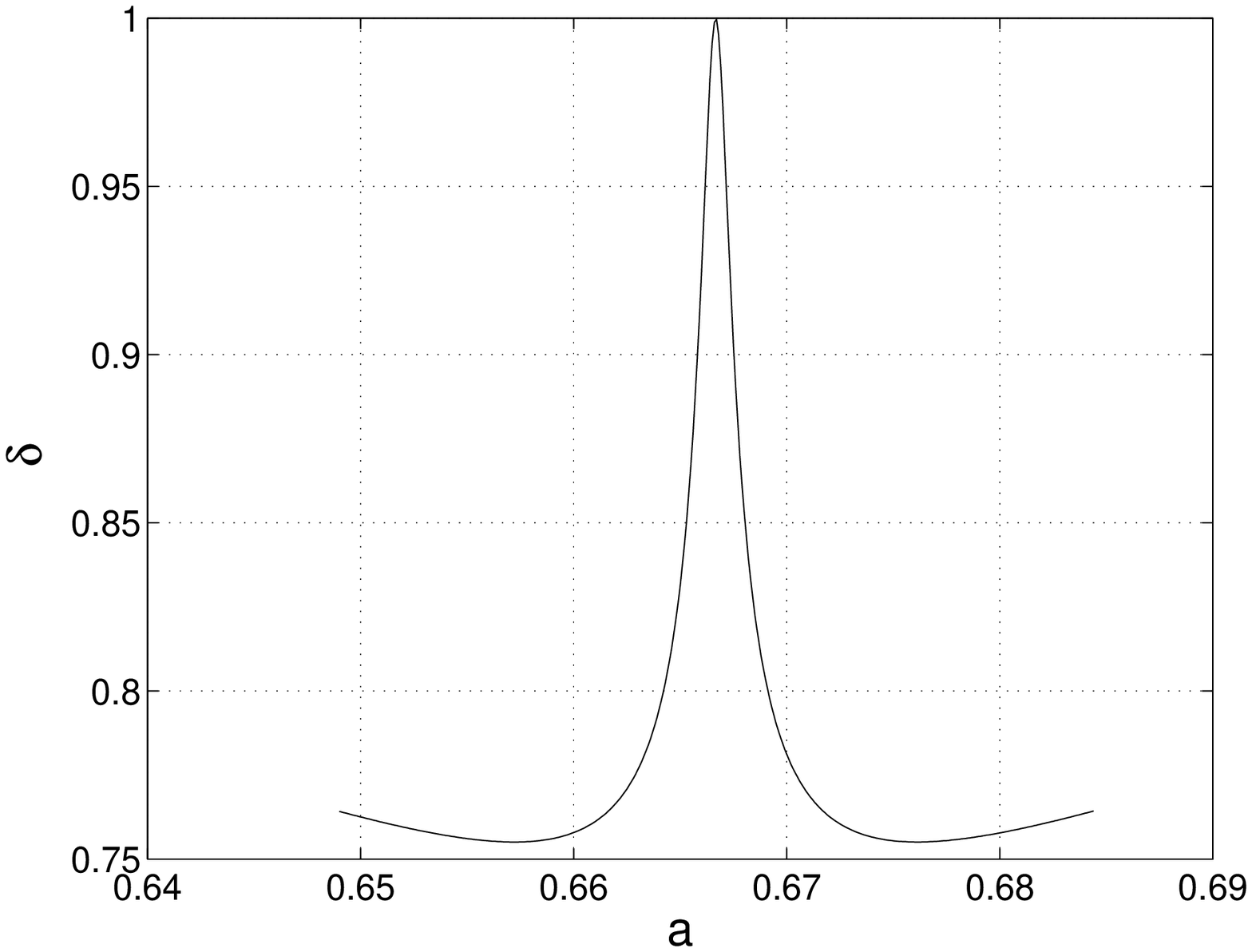,width=7.5cm}
\end{minipage}
\begin{minipage}[tr]{7.5cm}
\psfig{file=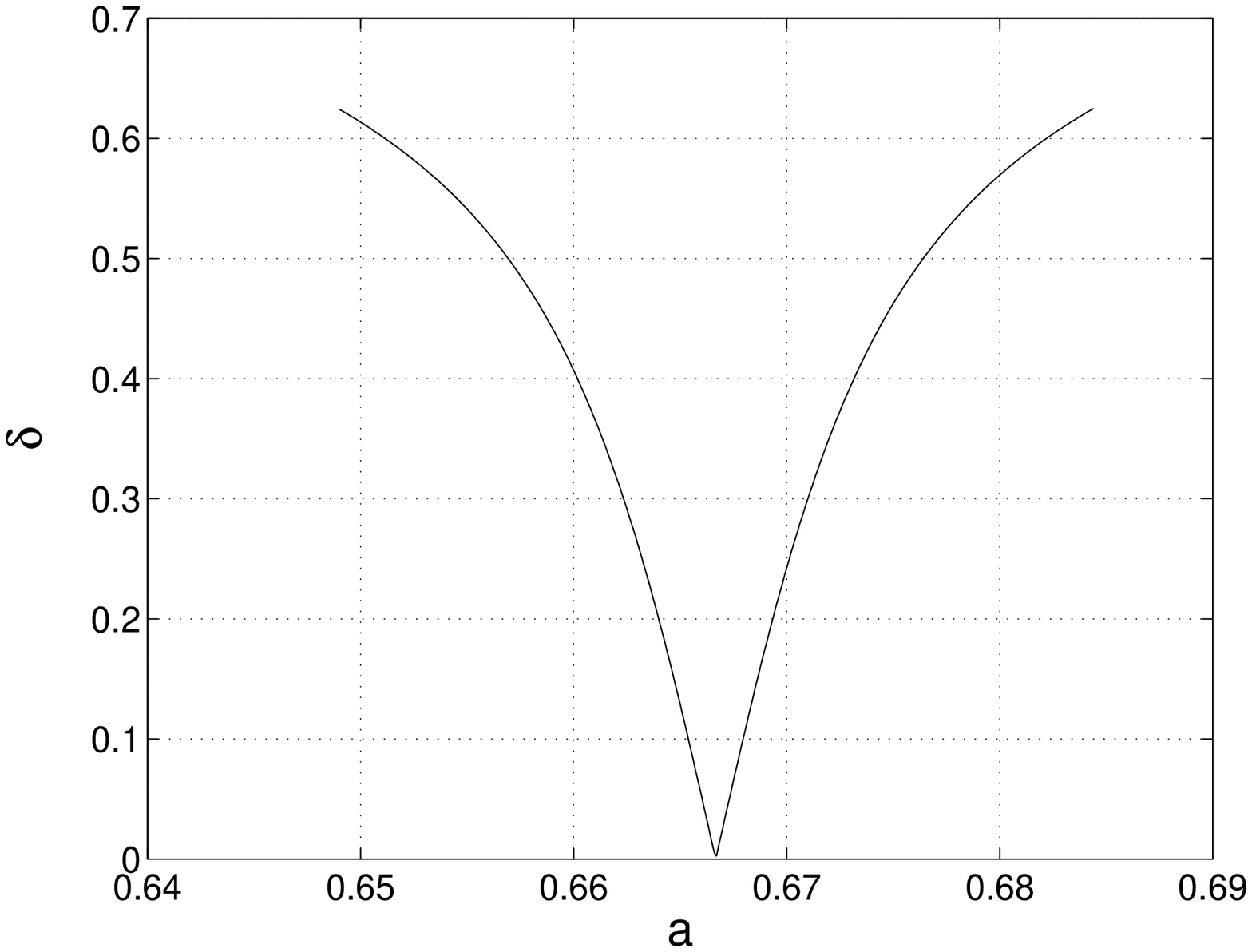,width=7.5cm}
\end{minipage}
\hspace*{.5cm}
\begin{minipage}[bl]{7.5cm}
\psfig{file=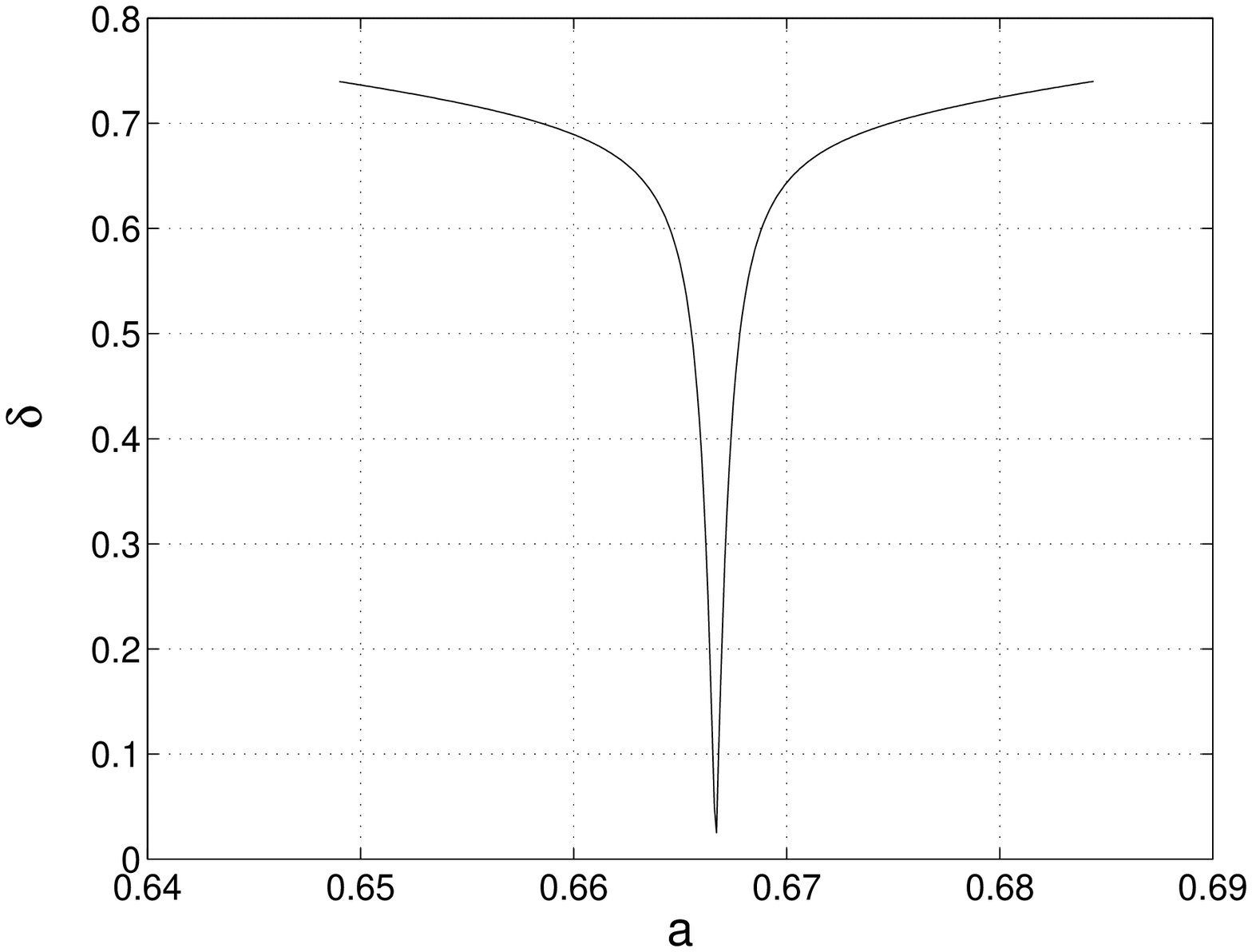,width=7.5cm}
\end{minipage}
\hspace*{.6cm}
\begin{minipage}[br]{7.5cm}
\psfig{file=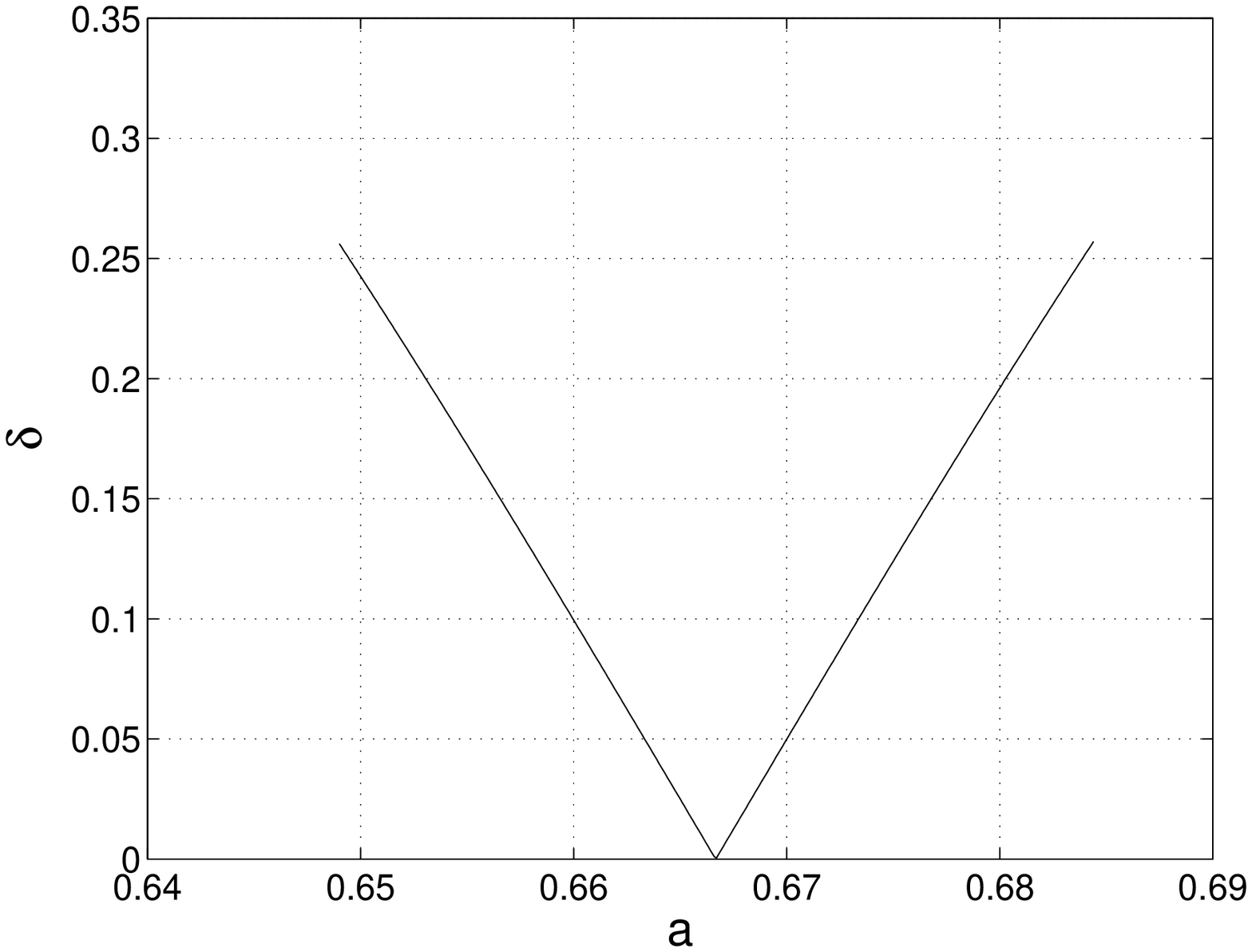,width=7.5cm}
\end{minipage}
\vspace*{.5cm}
\caption{The differences  $\delta= |b_{ii}|^2 - |b_{ij\ne i}|^2$ 
 as a function of the tuning parameter $a$.
 $e_1=1-a/2; \; e_2=a $ and $\omega = 0.05$. The 
$\gamma_1 /2 $  are 1.010  (top left), 0.990  (bottom left), 
0.90  (top right), 0  (bottom right); $\gamma _2 = 1.1 \cdot \gamma _1$.
Note the different scales in the  different figures.
}
\label{fig:avoi1}
\end{figure}

\begin{figure}
\begin{minipage}[tl]{7.5cm}
\psfig{file=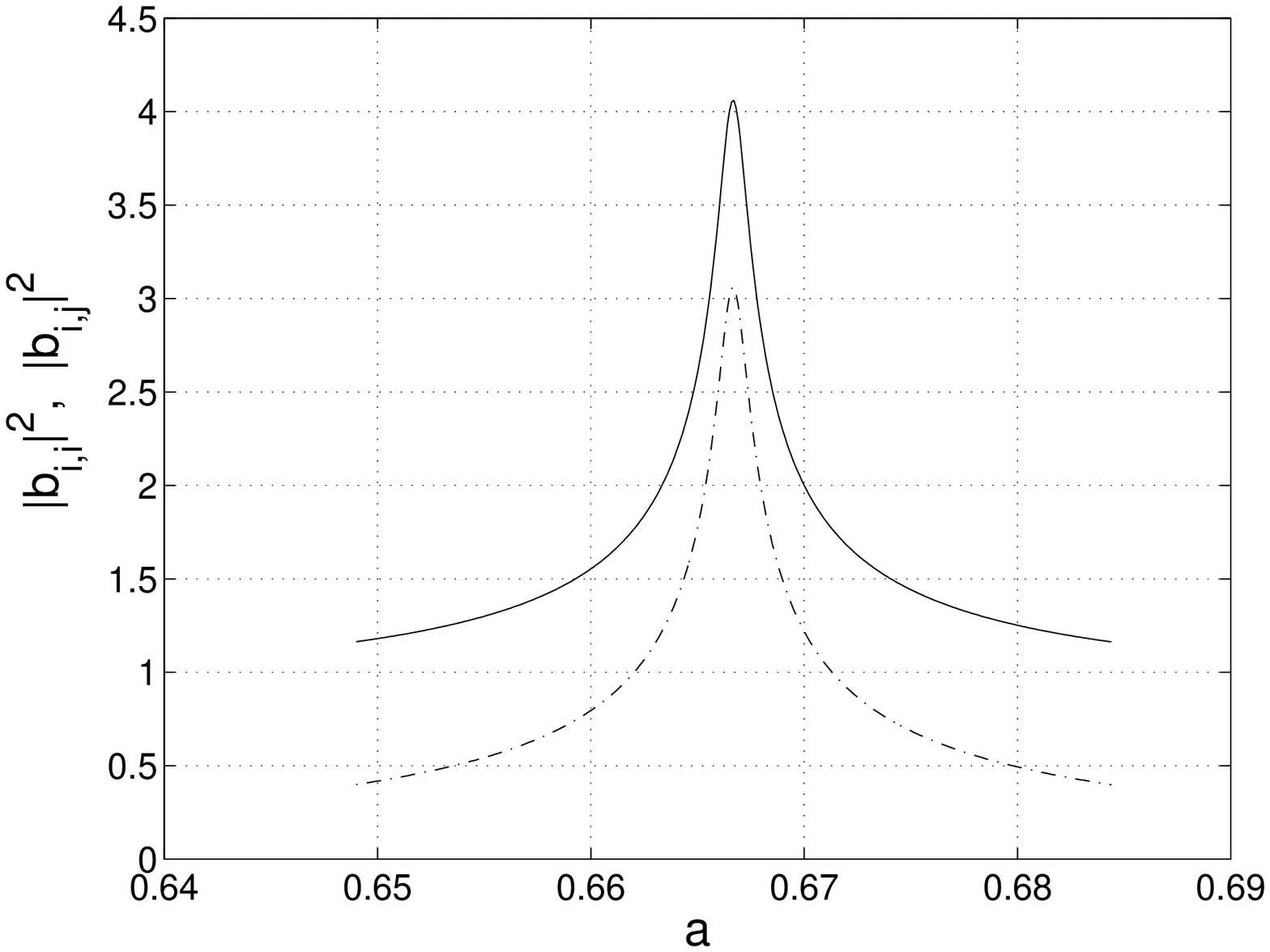,width=7.5cm}
\end{minipage}
\begin{minipage}[tr]{7.5cm}
\psfig{file=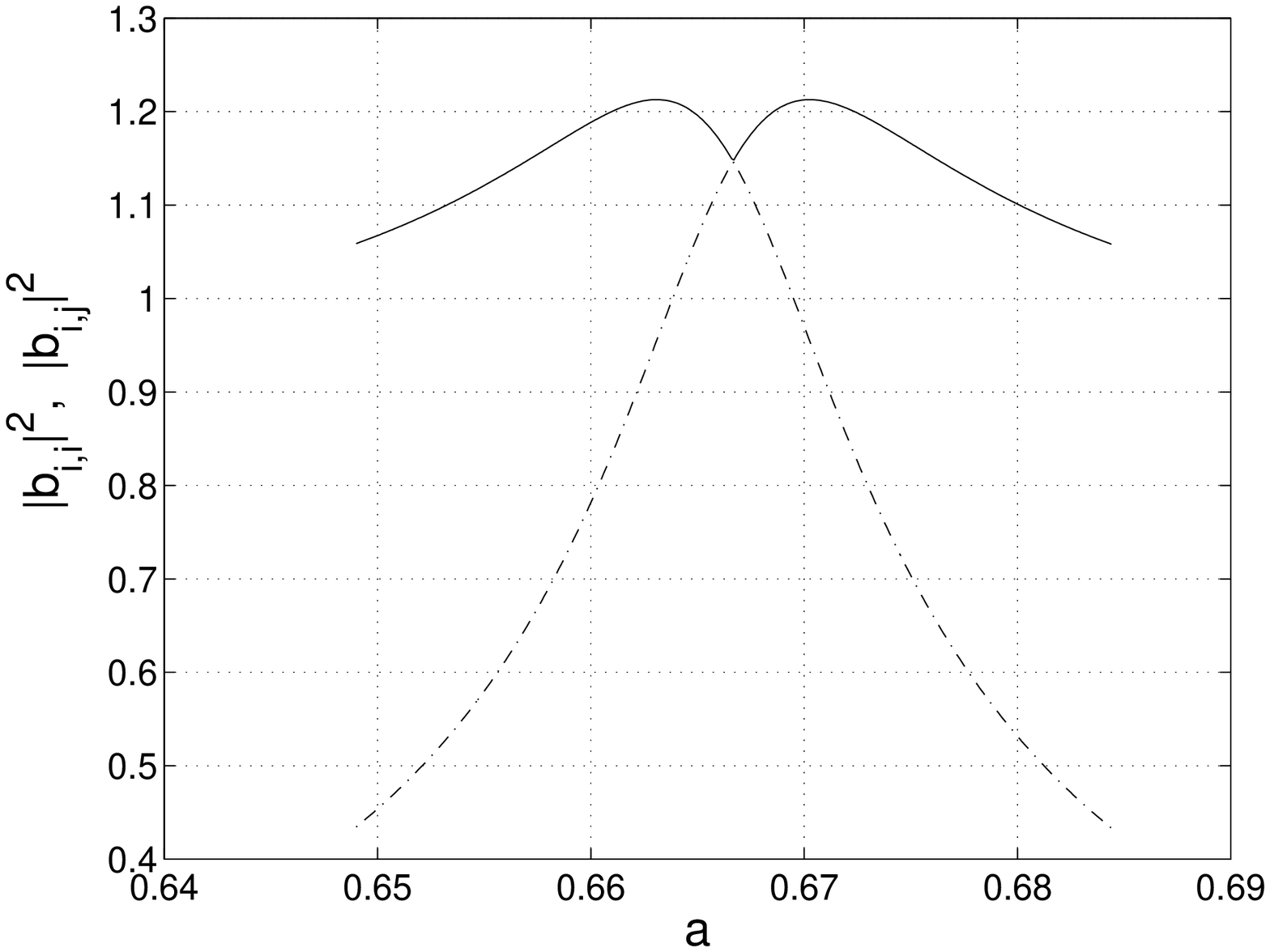,width=7.5cm}
\end{minipage}
\hspace*{.5cm}
\begin{minipage}[bl]{7.5cm}
\psfig{file=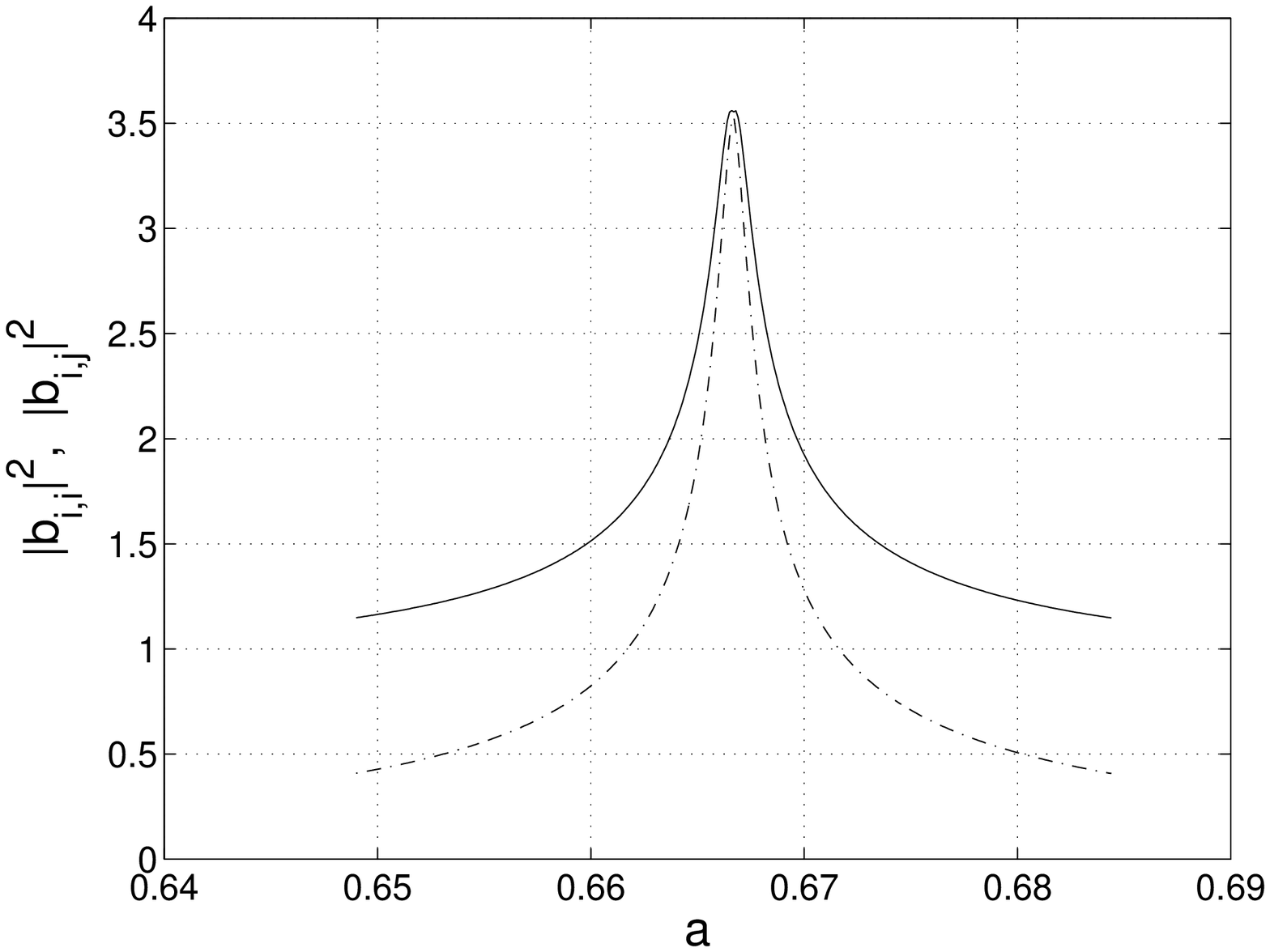,width=7.5cm}
\end{minipage}
\hspace*{.6cm}
\begin{minipage}[br]{7.5cm}
\psfig{file=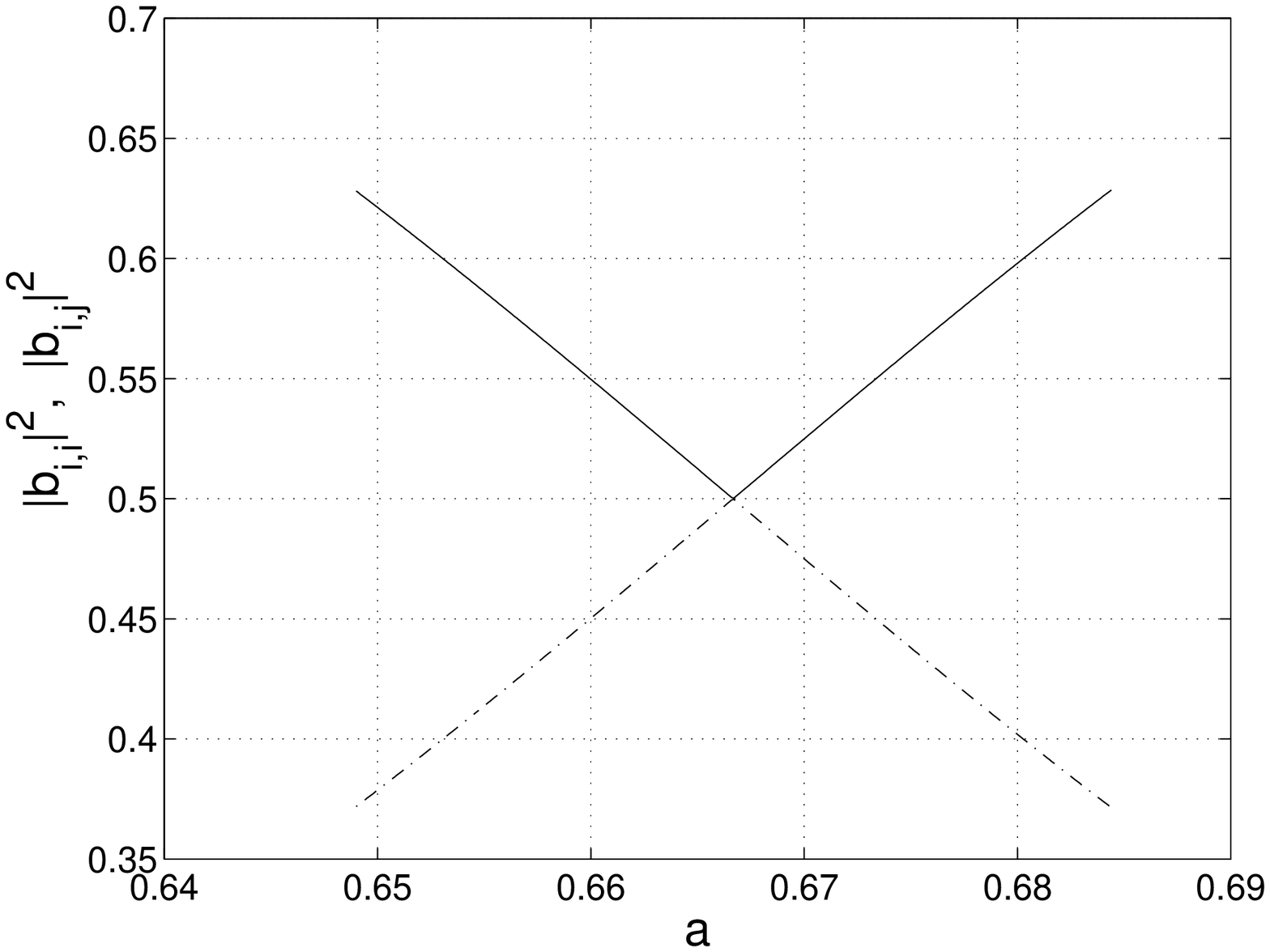,width=7.5cm}
\end{minipage}
\vspace*{.5cm}
\caption{
The  $|b_{ii}|^2$ (full lines)  and $ |b_{ij\ne i}|^2$ (dash-dotted lines) 
 as a function of the tuning parameter $a$.
 $e_1=1-a/2; \; e_2=a $ and $\omega = 0.05$. The $\gamma_1 /2 $ are 
the same  as in 
figure \ref{fig:avoi1}: 1.010  (top left), 0.990  (bottom left), 
0.90  (top right), 0  (bottom right); $\gamma _2 = 1.1 \cdot \gamma _1$.
Note the different scales in the  different figures.
}
\label{fig:avoi2}
\end{figure}

 At the double pole of the $S$ matrix,  not only 
 $\Gamma_+ = \Gamma_-$ but also $E_+ = E_-$
 for $a^{\rm cr}$. Here, the real and imaginary parts of
all components of the wavefunctions increase boundless and 
$\Phi_i \to \pm\; i \,
 \Phi_{j\ne i}$ \cite{marost}. Avoided level crossings are characterized by
$E_+ \ne E_-$ at  $a^{\rm cr}$. In these cases, 
the increase of the components of the wave functions at  $a^{\rm cr}$
is reduced due to interferences. The interferences 
can be seen, e.g., from the differences $\delta =
|b_{ii}|^2 - |b_{ij\ne i}|^2$ which
jump from 1 to 0 at the  branch point (see  figure
 \ref{fig:avoi1}). This jump is related to the normalization
 condition (\ref{eq:orth}) which is fulfilled also at the double pole of the 
$S$ matrix. The jump makes possible the exchange of the wave functions.
The value  $\delta = 0$ remains when $(\gamma_1 /2 - \gamma_2 /2)^2
 < 4 \; \omega^2$, i.e. also for discrete states.

\begin{figure}
\begin{minipage}[tl]{7.5cm}
\psfig{file=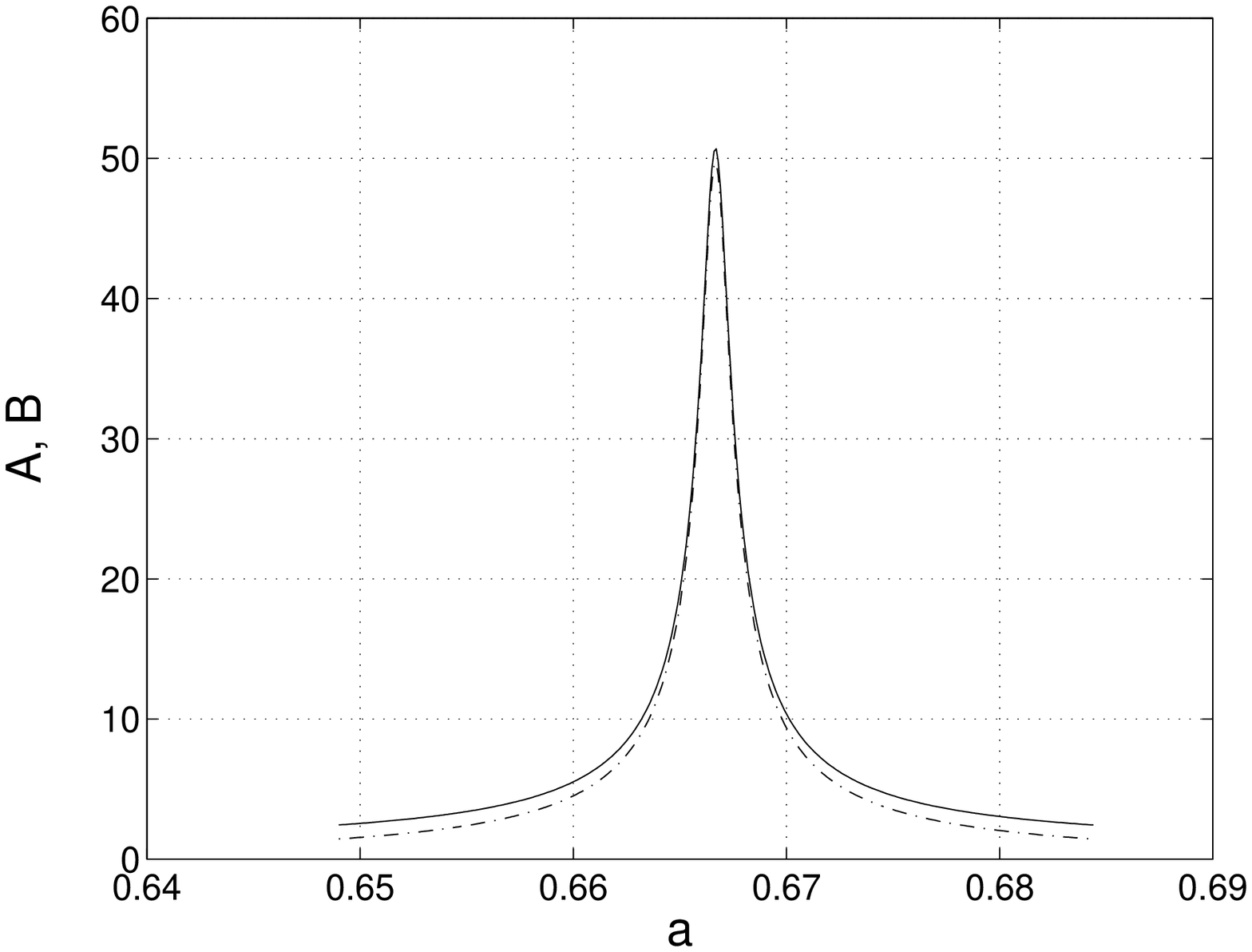,width=7.5cm}
\end{minipage}
\begin{minipage}[tr]{7.5cm}
\psfig{file=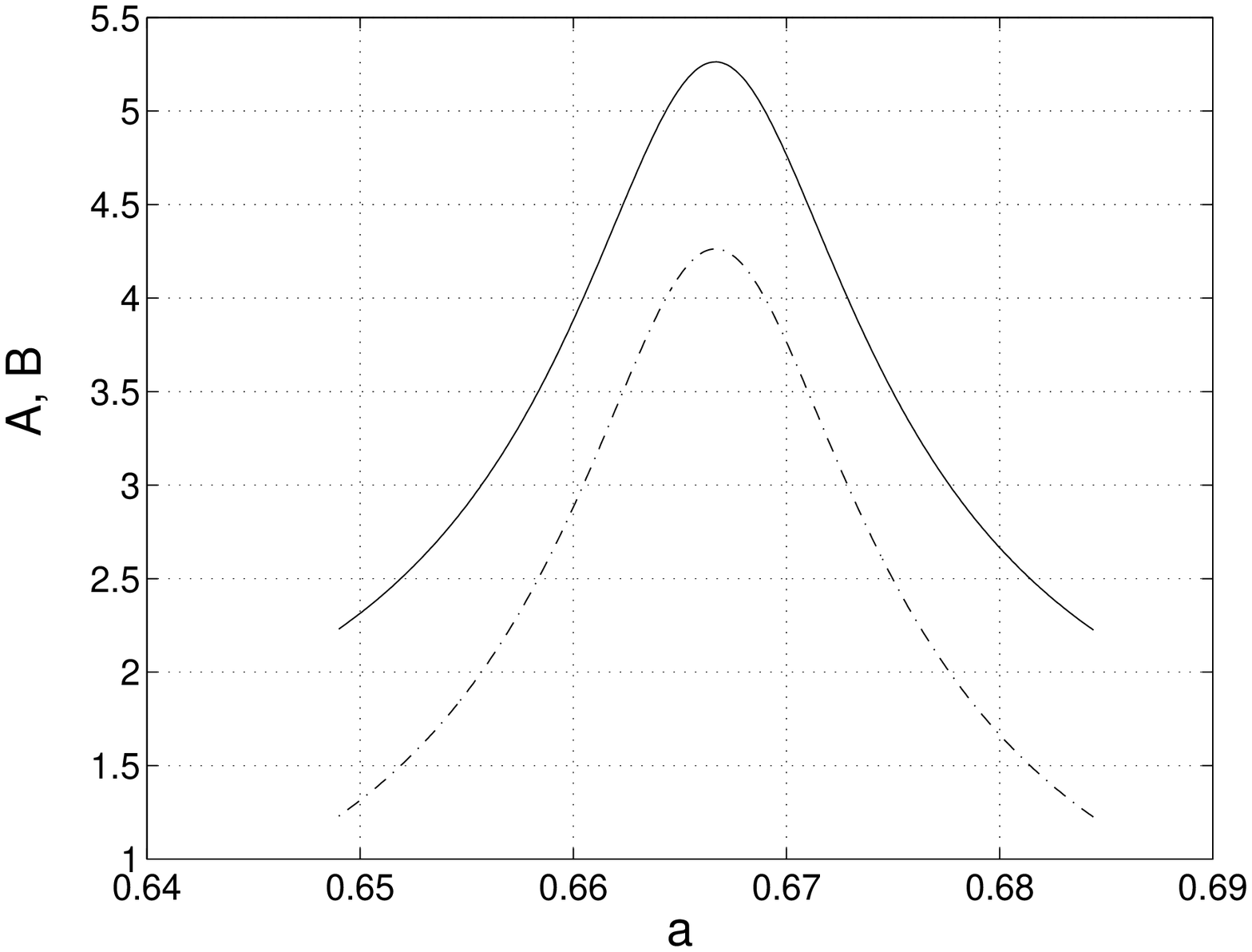,width=7.5cm}
\end{minipage}
\hspace*{.5cm}
\begin{minipage}[bl]{7.5cm}
\psfig{file=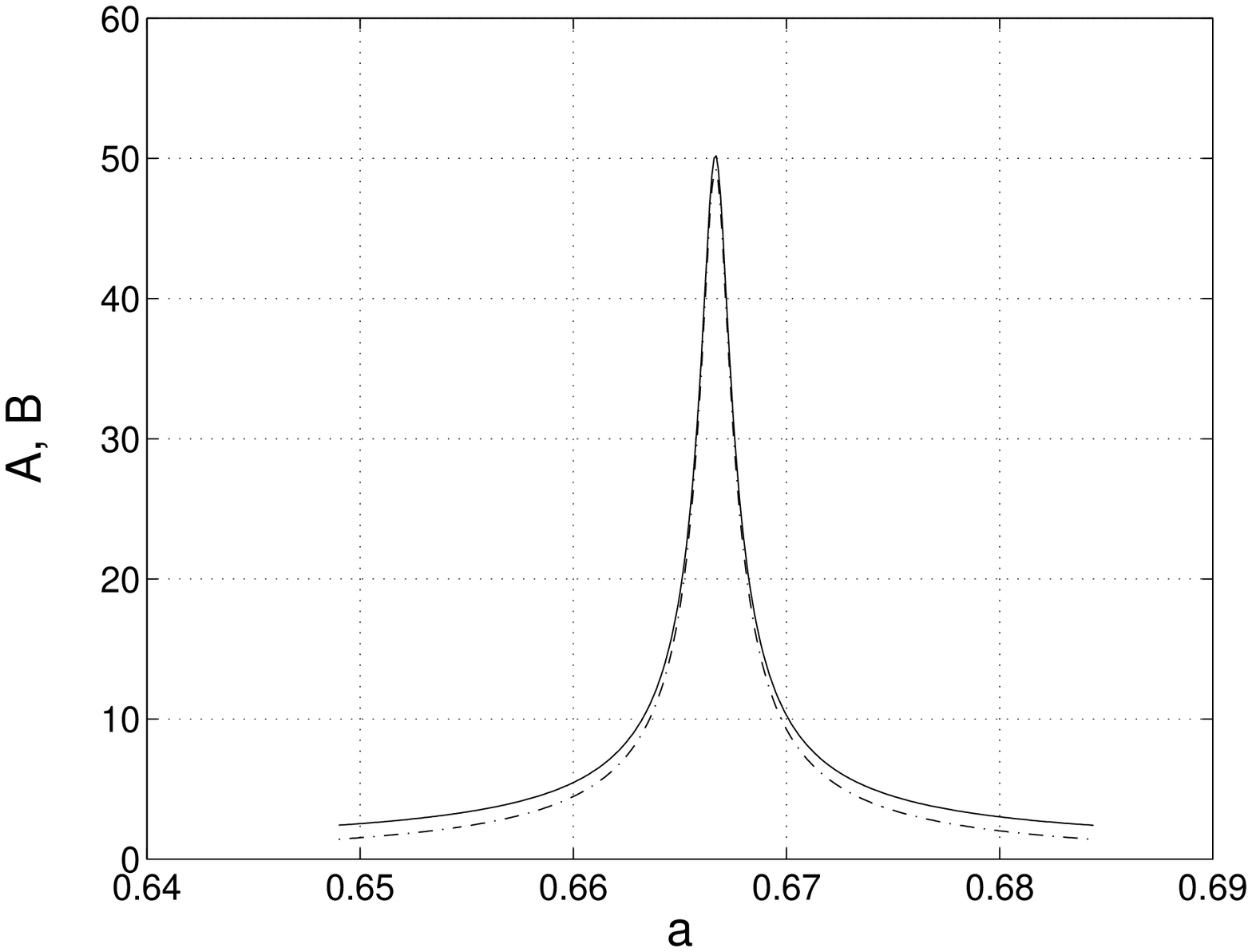,width=7.5cm}
\end{minipage}
\hspace*{.6cm}
\begin{minipage}[br]{7.5cm}
\psfig{file=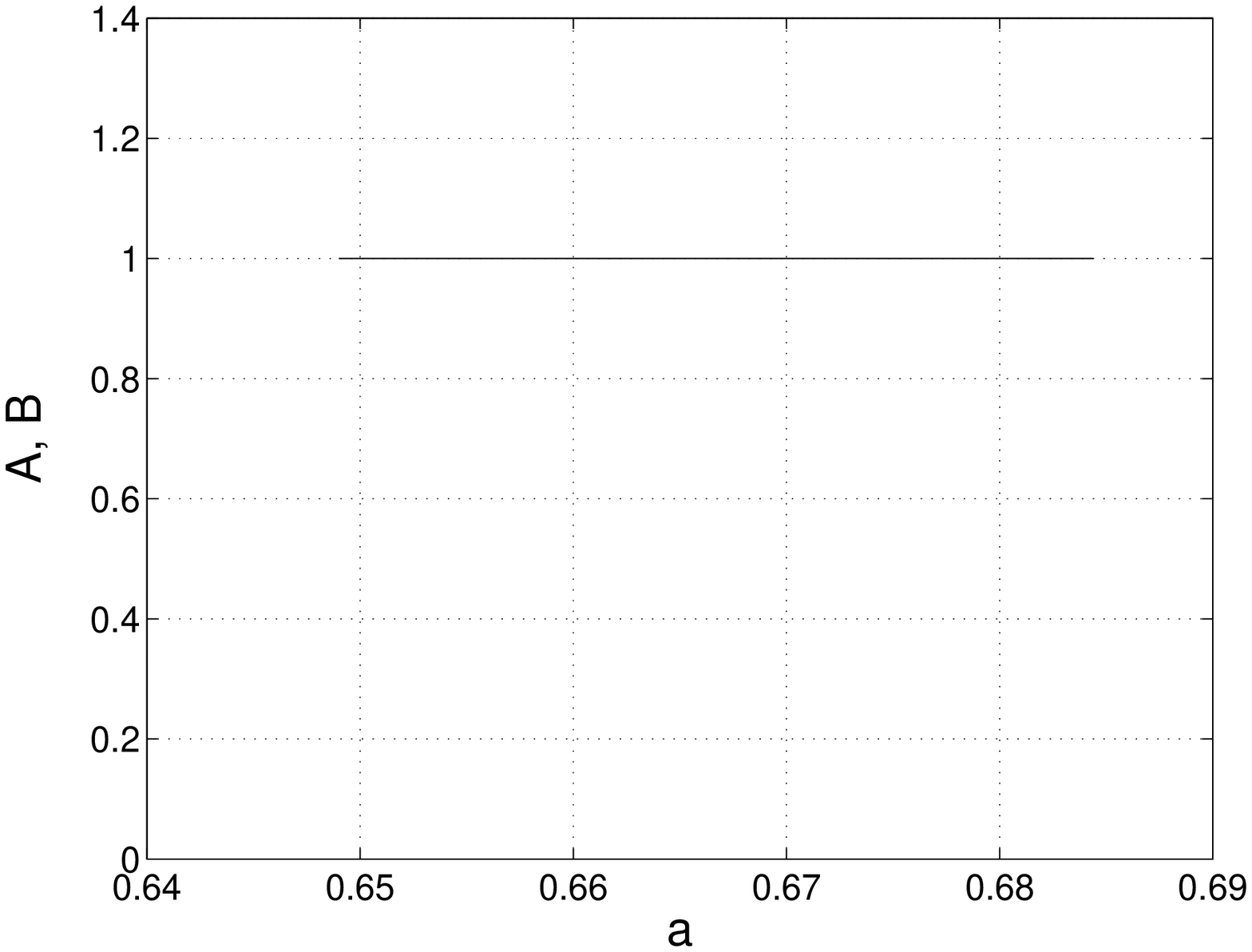,width=7.5cm}
\end{minipage}
\vspace*{.5cm}
\caption{
The  $A$ (full lines)  and $ B$ (dash-dotted lines) defined in equation
(\ref{eq:biorth})
 as a function of the tuning parameter $a$.
  $e_1=1-a/2; \; e_2=a $ and $\omega = 0.05$. The 
$\gamma_1 /2 $  are the same as in 
figure \ref{fig:avoi1}:  1.010  (top left), 0.990  (bottom left), 
0.90  (top right), 0  (bottom right); $\gamma _2 = 1.1 \cdot \gamma _1$.
Note the different scales in the  different figures.
}
\label{fig:avoi3}
\end{figure}

The bi-orthogonality (\ref{eq:biorth}) of the wave functions characterizes the
avoided crossing of resonance states. It increases boundless at the double
pole of the $S$ matrix and vanishes in the case of an 
avoided crossing of discrete states (figure \ref{fig:avoi3}).

The results of all the calculations presented in the present paper show very
clearly that, from a mathematical point of view, the branch points in the 
complex plane continue analytically into the function space of discrete
states. Their influence on the properties of discrete states cannot be
neglected.

\begin{figure}
\hspace{-1.8cm}
\begin{minipage}[tl]{7.5cm}
\psfig{file=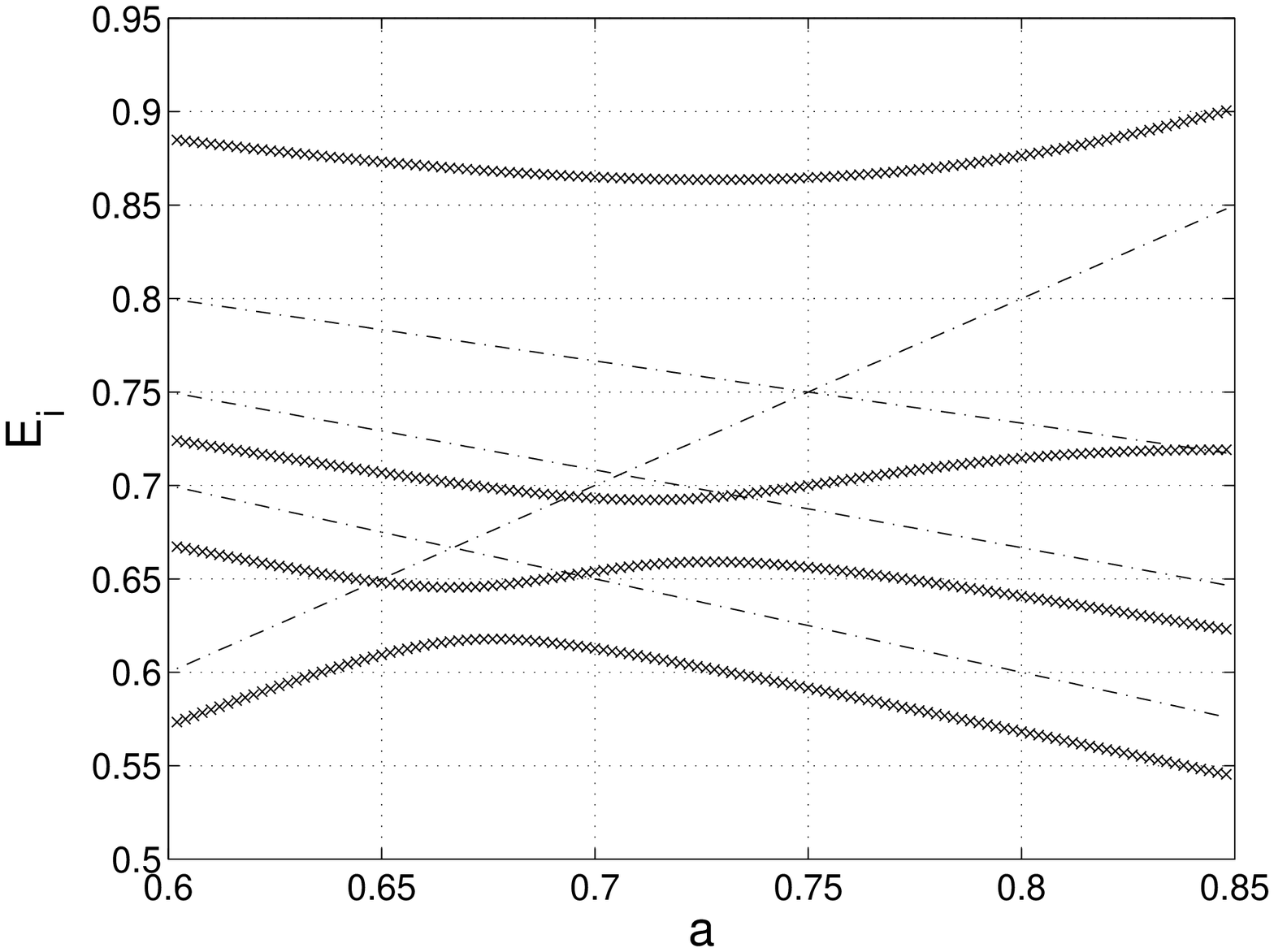,width=7.5cm}
\end{minipage}
\begin{minipage}[tr]{7.5cm}
\psfig{file=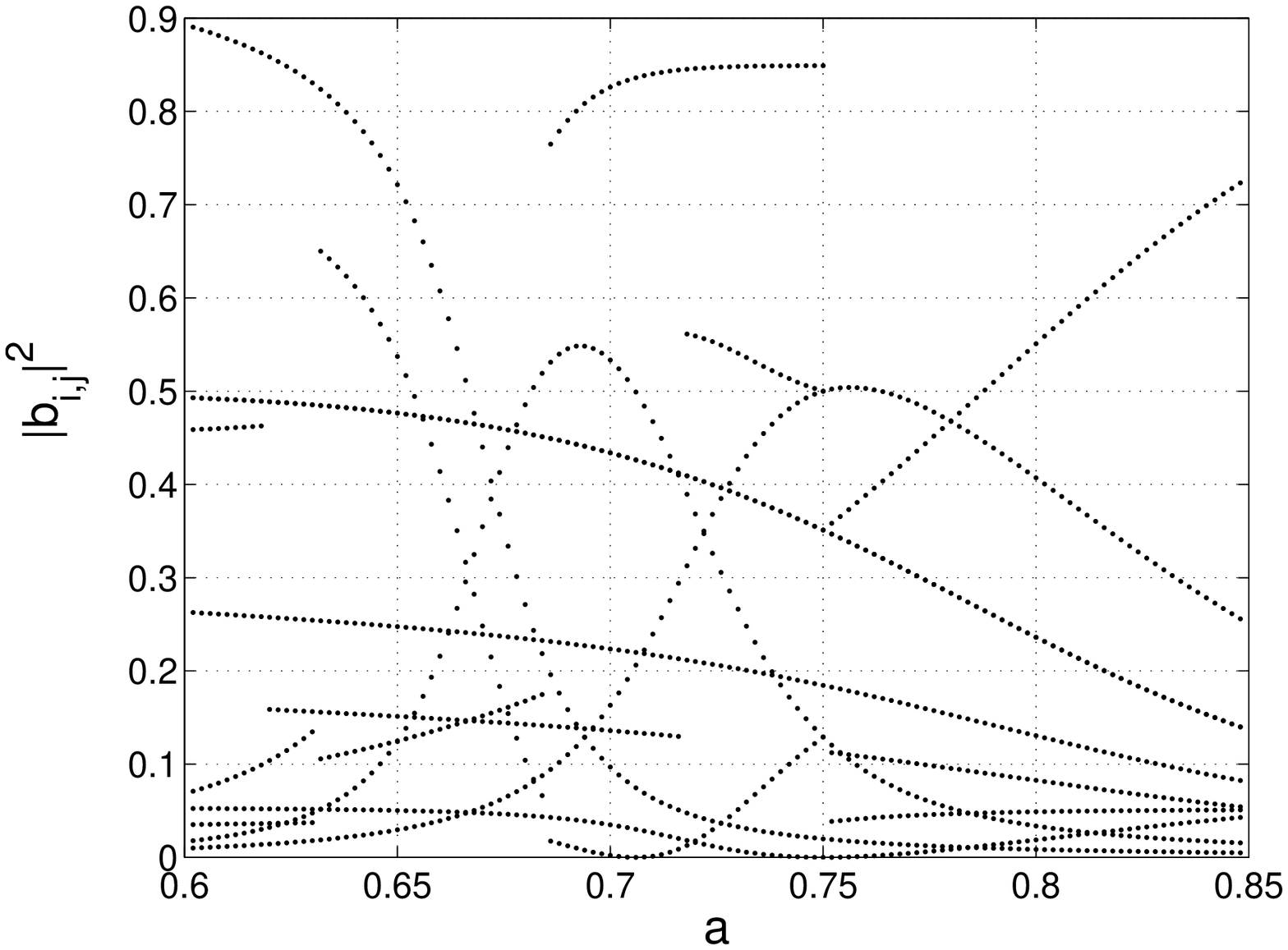,width=7.5cm}
\end{minipage}
\hspace{1cm}
\begin{minipage}[ml]{7.5cm}
\psfig{file=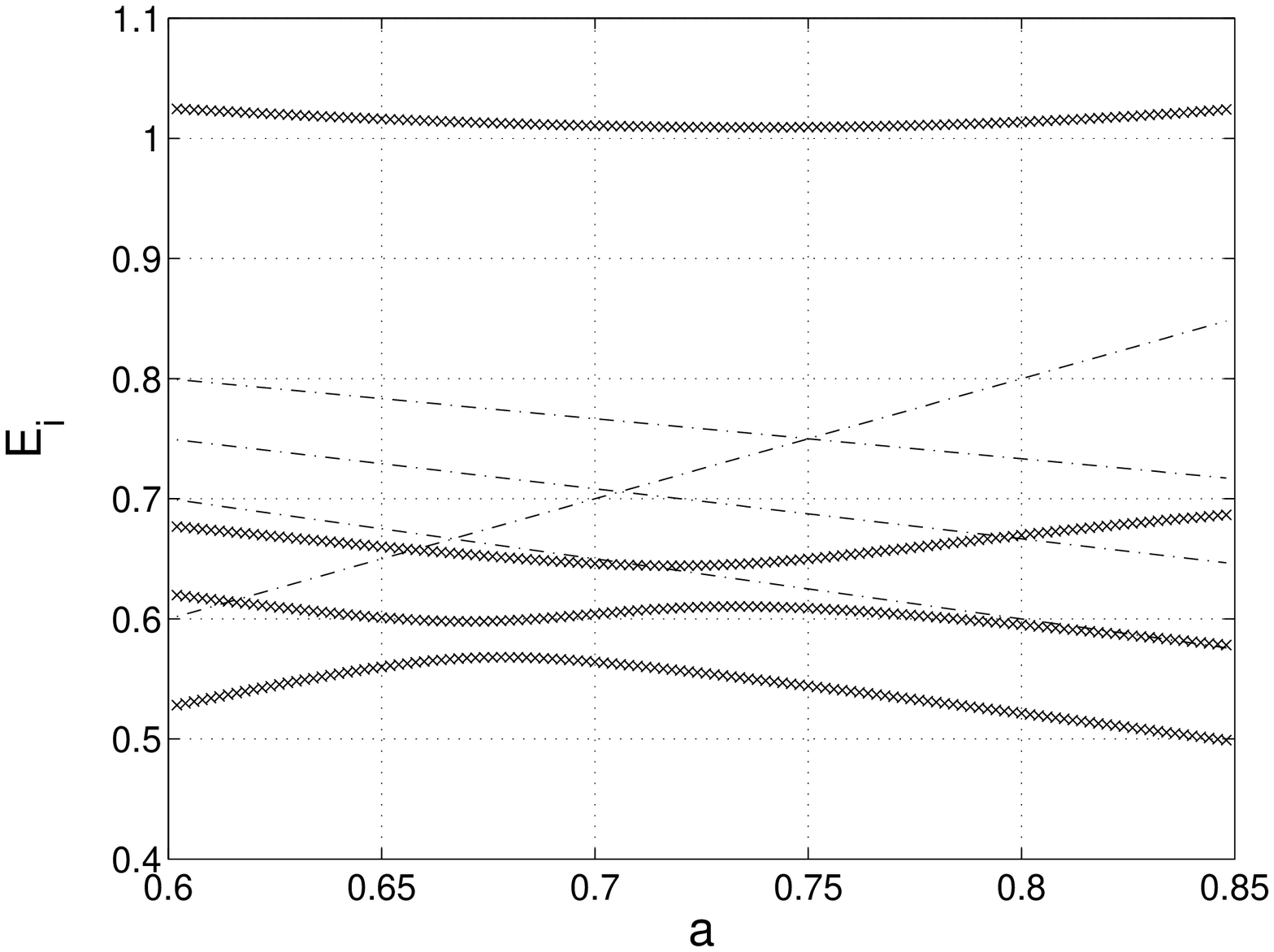,width=7.5cm}
\end{minipage}
\begin{minipage}[mr]{7.5cm}
\psfig{file=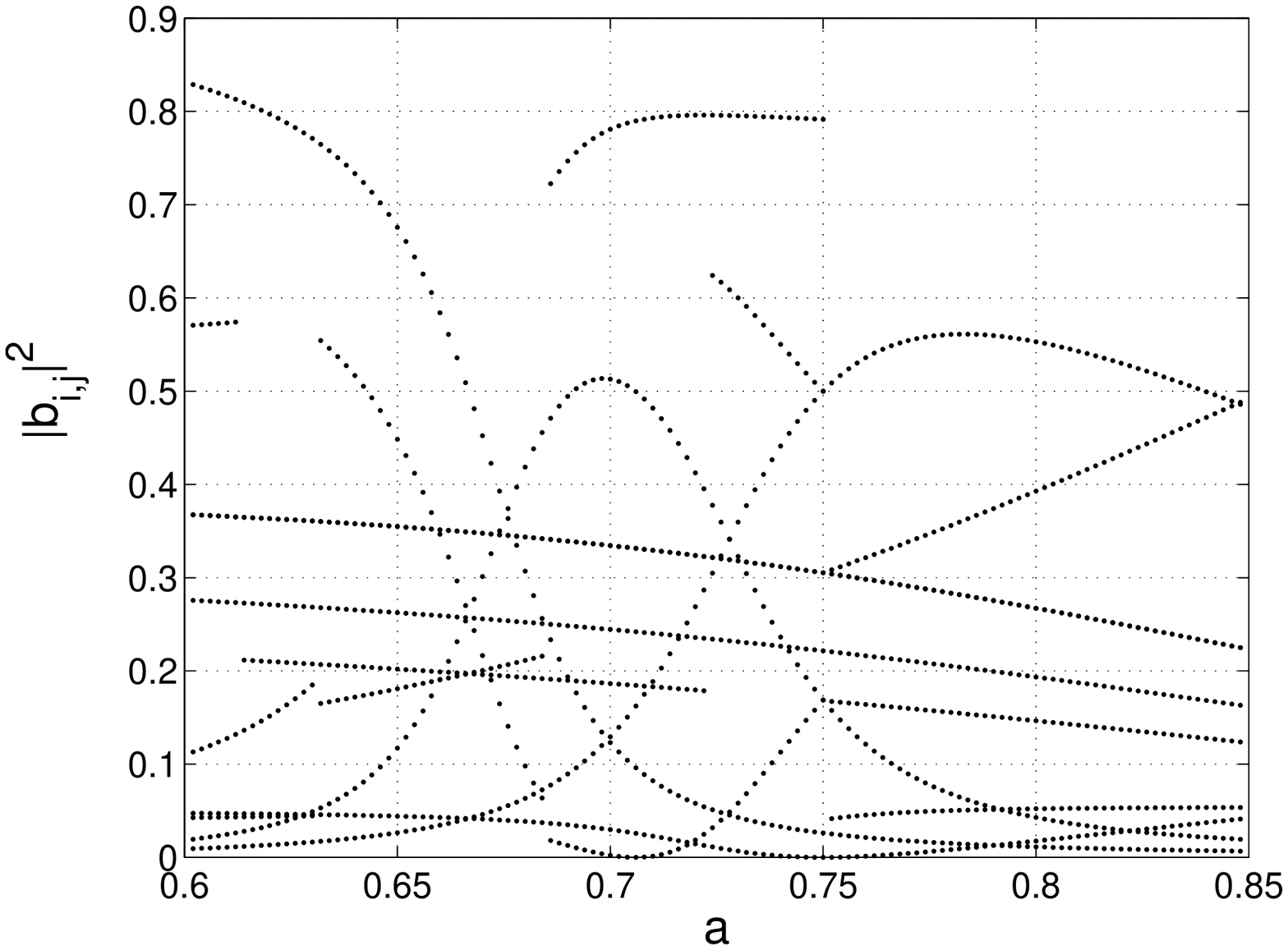,width=7.5cm}
\end{minipage}
\begin{minipage}[bl]{7.5cm}
\psfig{file=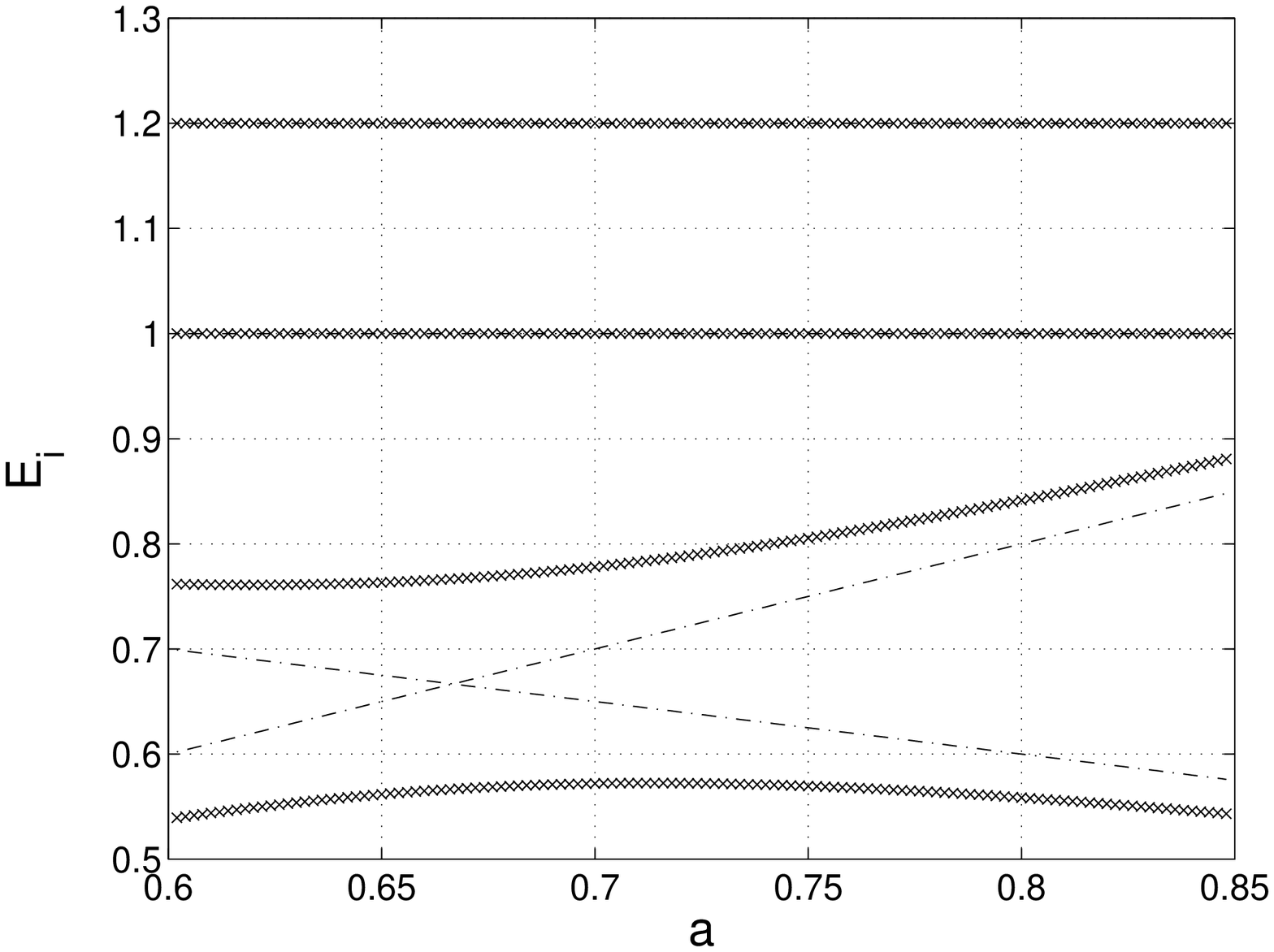,width=7.5cm}
\end{minipage}
\hspace*{1cm}
\begin{minipage}[br]{7.5cm}
\psfig{file=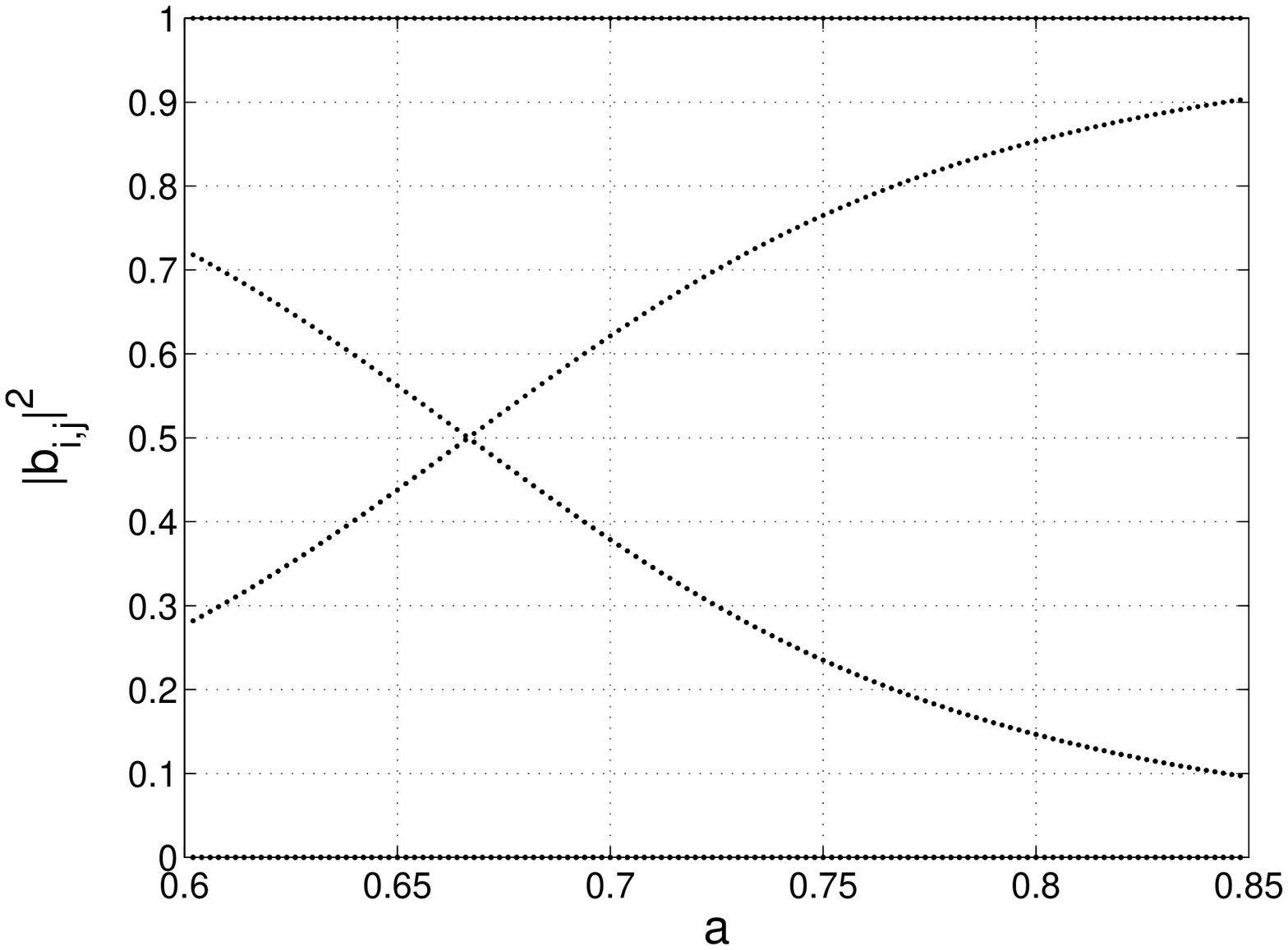,width=7.5cm}
\end{minipage}
\vspace*{.5cm}
\caption{The energies $E_i$ (left) and mixing coefficients $|b_{ij}|^2$
(right) of four discrete states ($\gamma_i = 0$ for $i= 1, ..., 4$)
obtained from ${\cal H}^{(4)}$, equation (\ref{eq:matr4}), as a function of the
tuning parameter $a$. Top and middle: $e_1=1-a/3; \; e_2=1- 5 a / 12 ; \; 
e_3=1-a/2; \;  e_4=a $; 
$~\omega = 0.05$ (top) and 0.1 (middle) for all non-diagonal matrix elements.
Bottom:  the same as above but 
 $e_1=1; \; e_2=1.2$;  $\omega =0$ for the coupling between
the states  $i=1,2$ and $j\ne i$,  $\; \omega =0.1$ for the coupling between
$i=3,4$ and $j=4,3$. In this case,  $|b_{ii}|^2  \ge |b_{ij\ne i}|^2$
(bottom right) as in figure \ref{fig:avoi2} (bottom right). 
The dash-dotted lines  (left)  show  $E_i$  for $\omega = 0$. 
The  states $i$ and $j$ are exchanged at some values $a$.
}
\label{fig:four}
\end{figure}

Another result of the present study is the influence of the branch 
points  in the complex plane  
onto the purity of the wave functions $\Phi_\pm$. 
At  $a^{\rm cr}$, the wave
functions are not only exchanged but become mixed. The mixing occurs not only
at the critical point  $a^{\rm cr}$ but in a certain region  around
$a^{\rm cr}$ when the crossing  is avoided. 
This fact is important at high level density where, as a rule, 
an avoided crossing with another level appears before $\Phi_i \to
\Phi^0_{j}$ is reached. As a result, all the wave functions of closely-lying
states are strongly mixed, i.e. the states are strongly
correlated at high level density (see figure  \ref{fig:four}). 

The strong mixing of the wave functions of a quantum system at high level
density means that the information on the individual properties of the  
states $\Phi_i^0$ is lost.  While the exchange of the wave
functions itself is of no interest for a statistical consideration of the
states, the accompanying correlation of the states is decisive for the 
statistics. This fact is discussed also in \cite{haake}. As shown in 
the present paper, the correlation of the states 
may be traced back to the existence of branch points in the
complex plane. The number of branch points in the complex plane 
is large at high level density. Therefore, the discrete states of
quantum systems at high level density 
lose any information on their individual properties. That means, they
contain minimum information. 

Thus, there is an information exchange between a (closed) quantum system 
and the continuum due to the analyticity of the wave functions. The branch 
points in the complex plane are {\it hidden crossings}, indeed. They play an
important role not only in atoms, as supposed in \cite{solov1,solov2}, 
but determine the 
properties of all quantum systems at high level density. 

The loss of information on the properties of discrete states of a quantum 
system under the influence of branch points in the complex plane
 is in complete analogy to 
the conclusion drawn from the statistical properties of the city transport 
in Cuernavaca (Mexico)  \cite{seba}. Also in this case,
an information exchange with
the environment leads to an information loss in the system.

It follows further that the statistical properties of quantum systems at 
high  level density are different from those at low level density. 
States at the border of the spectrum are almost not influenced by 
branch points in the complex plane since their number is  small.
There are almost no states  which
could cross or  avoid crossing with others states. 
The properties of these states are
expected therefore to show more individual features than those at high level
density.  In other words, the information on the individual properties of the
states $\Phi_i^0$ at the border of the spectrum  
is kept to a great deal in contrast to
that on the states in the center of the spectrum.

Summarizing the results, it can be stated the following.
Branch points in the complex plane cause  an exchange of the wave
functions and, above all,
create  correlations between the states of a
quantum system at high level density even if the system is closed and the 
states are discrete. These correlations are equivalent to a
loss of information on the individual basic states. 
This result may explain the similarity between the statistical properties
of the city transport in Cuernavaca (Mexico) and  those of a quantum system
at high level density.

\vspace*{.8cm}
\noindent
{\bf Acknowledgment}: Valuable discussions with E. Persson, J.M. Rost, 
 P. ${\rm \check S}$eba and E.A. Solov'ev
are gratefully acknowledged. I am indebted to M. Sieber
 for a critical reading of the manuscript.

\newpage

\end{document}